\documentstyle[preprint,prd,aps,eqsecnum,amssymb,epsf]{revtex}

\tighten

\begin{document}
\draft
\preprint{DAMTP-R96/11}
\title{Classical and Quantum Initial Value Problems \\ for Models of
Chronology Violation}
\author{C.J. Fewster\thanks{Electronic address: cjf3@unix.york.ac.uk} and
A. Higuchi\thanks{Electronic address: ah28@unix.york.ac.uk}}
\address{Institut f\"{u}r theoretische Physik, Universit\"{a}t Bern, 
Sidlerstrasse 5, CH-3012 Bern, Switzerland \\
and Department of Mathematics, University of York, \\
Heslington, York YO1 5DD, United Kingdom\footnote{Current address of
both CJF and AH}}
\author{C.G. Wells\thanks{Electronic address: C.G.Wells@damtp.cam.ac.uk}}
\address{Department of Applied Mathematics and Theoretical Physics,
University of Cambridge, \\ Silver St., Cambridge CB3 9EW, United
Kingdom} 
\date{\today}
\maketitle

\begin{abstract}

  We study the classical and quantum theory of a class of nonlinear 
differential equations on chronology violating spacetime models in which space 
consists of finitely many discrete points. Classically, in the linear and 
weakly nonlinear regimes (for generic choices of parameters) we prove existence
and uniqueness of solutions corresponding to initial data specified before the 
dischronal region; however, uniqueness (but not existence) fails in the 
strongly coupled regime. The evolution preserves the symplectic structure. 

  The quantum theory is approached via the {\em quantum initial value problem} 
(QIVP); that is, by seeking operator-valued solutions to the equation of 
motion with initial data representing the canonical (anti)commutation 
relations. Using normal operator ordering, we construct solutions to the QIVP 
for both Bose and Fermi statistics (again for generic choice of parameters) 
and prove that these solutions are unique. For models with two spatial points,
the resulting evolution is unitary; however, for a more general model the 
evolution fails to preserve the (anti)commutation relations and is therefore 
nonunitary. We show that this nonunitary evolution {\em cannot} be described 
using a superscattering operator with the usual properties. 

  We present numerical evidence to show that the bosonic quantum theory can 
pick out a unique classical limit for certain ranges of the coupling strength,
even when there are many classical solutions. We show that the quantum theory 
depends strongly on the choice of operator ordering. 

  In addition, we show that our results differ from those obtained using the 
``self-consistent path integral''. It follows that the path integral evolution
does not correspond to a solution of the equation of motion. 

\end{abstract}


\narrowtext

\section{Introduction}

Spacetimes containing closed timelike curves (CTC's) provide an
intriguing environment for the formulation of both classical and quantum   
physics. Because the present is influenced by both the past
and the future, neither existence nor uniqueness is guaranteed {\em a
priori} for solutions to initial value problems for particles and fields
on such spacetimes; these issues underlie
many of the apparent paradoxes associated with time travel. In this
paper, we attempt to gain insight into the initial value problem for a
class of nonlinear differential equations (which may be regarded as toy
field theories) on ``spacetimes'' of a type introduced by
Politzer~\cite{Pol2}. These spacetimes are defined by taking the
Cartesian product of a
number of discrete points (representing space) with the real line (time) 
and then imposing certain identifications to introduce CTC's. The simple 
nature of these models removes many technical problems and allows us to
pursue the analysis to its end. 

Previous studies of classical initial value problems on chronology
violating spacetimes have mostly focussed on linear
fields~\cite{FM1,FM2,FMNEKTY,Pol,GPPT} and billiard ball
models~\cite{FMNEKTY,EKT,Novikov,RamaSen}. Deutsch~\cite{Deut} has also
studied examples of classical computational networks with chronology
violating components. To the best of
our knowledge there have been no studies of nonlinear fields, except
insofar as the billiard ball models may be regarded as providing insight
into the strongly nonlinear limit. For linear fields, it turns out that
one can formulate a well posed initial value problem under certain
conditions. Friedman and Morris~\cite{FM1,FM2} have rigorously proved 
existence and partial uniqueness results for massless fields propagating
on a class of smooth static wormhole spacetimes with data specified at
past null infinity. In addition, they have conjectured that the initial
value problem
is well posed for asymptotically flat spacetimes with a compact
nonchronal region whose past and future regions are globally hyperbolic
whenever the problem is well posed in the geometric optics limit. It is
much easier to prove existence and uniqueness for linear fields on
certain non-smooth chronology violating spacetimes~\cite{Pol,GPPT}. 

In the billiard ball case, Echeverria {\em et al.}~\cite{EKT} and (for
similar and more elaborate systems)
Novikov~\cite{Novikov} have shown that the initial value problem is
often ill posed in the sense that the evolution is nonunique; moreover,
Rama and Sen~\cite{RamaSen} have given similar examples in which there
appears to be no global self-consistent solution for certain initial
data.  

One would expect that nonlinear fields should interpolate between the
behaviour exhibited by linear fields on one hand and billiard balls on
the other. In this paper, we will show that this is indeed the case for 
our class of models: we prove that the initial value problem is well
posed for arbitrary data specified before the dischronal region
in both the linear and weakly nonlinear case, but uniqueness
(though not existence) fails in the strongly nonlinear regime. In
addition to these analytic results, we give an explicit example to
demonstrate the lack of uniqueness for a particular value of ``coupling
strength''. We also show that the evolution from the past of the
dischronal region to its future preserves the symplectic structure. 

The loss of uniqueness for interacting systems on chronology violating
spacetimes entails that classical physics loses its predictive
power. Various authors have expressed the hope that quantum dynamics on
such spacetimes might be better behaved than its classical counterpart,
with attention focussing on spacetimes possessing both initial and
final chronal regions. Friedman {\em et al.}~\cite{FPS2} considered
linear quantum fields and showed that,
provided the classical initial value problem is well posed, the quantum 
evolution between spacelike surfaces in the initial and final chronal
regions is unitary; a conclusion borne out by Boulware~\cite{Boul} in a 
Gott space example (see also~\cite{Pol2,Pol,GPPT} for related results).
However, the situation is very different for interacting fields. Both
Boulware~\cite{Boul} and Friedman {\em et al.}~\cite{FPS1} found that
the $S$-matrix between the initial and final chronal regions fails
to be unitary in perturbative $\lambda\phi^4$ theory. Politzer also
obtained similar perturbative results in quantum mechanics~\cite{Pol}
and also some nonperturbative results in exactly soluble
models~\cite{Pol2} in which nonunitarity also arises. It is also worth
pointing out that some interacting systems do have unitary quantum
theories~\cite{Pol}. 

The breakdown of unitarity raises many problems for the probability 
interpretation of quantum theory; in particular, ambiguities arise in
assigning probabilities to the outcomes of measurements conducted
before~\cite{FPS1} or spacelike separated from~\cite{Jacobson} the
nonchronal region. There have been various reactions to these problems.
Firstly, Hartle~\cite{Hartle} has discussed how 
nonunitary evolutions can be accommodated within the framework of 
generalised quantum mechanics, and a similar proposal has also been
advanced by Friedman {\em et al.}~\cite{FPS1}. A second approach has been
to ``repair'' the theory by modifying the evolution to yield a unitary
theory~\cite{Arley,FW}. Thirdly, Hawking~\cite{Hawk} has argued that one
should expect loss of quantum coherence in the presence of CTC's and that
the evolution should be specified by means of a superscattering operator
(i.e., a linear mapping from initial to final density matrices)
which moreover would not factorise into a unitary $S$-matrix and its
adjoint. On this view, the quantity computed using the usual rules for
the $S$-matrix is not the physically relevant quantity and its
nonunitarity is irrelevant. Rather, one should compute the matrix
elements of the superscattering operator. Deutsch~\cite{Deut} has also
advocated a density matrix formalism in the context of quantum
computational networks (see also~\cite{Pol2}). However, this
prescription turns out to be nonlinear in the initial density
matrix~\cite{Cass}. 

For the most part, the quantisation method employed in discussions of
chronology violation has been based on path integrals in which one sums
over all consistent trajectories or field configurations. In this paper,
we follow an operator approach based on the {\em quantum initial value
problem} (QIVP). Namely, we seek operator valued solutions to the
equation of motion and any consistency conditions arising from the CTC's,
with initial data specified before the dischronal region and forming a
representation of the canonical (anti)commutation relations. If there
exists a unique solution with this initial data, and the evolved data to
the future of the dischronal region also represents the commutation
relations, then we say that the QIVP is well posed. 

We will show that the QIVP is well posed for all linear models in our
class of interest with both Bose and Fermi statistics. The corresponding
quantum theory is unitary and agrees with that derived by path integral
methods. In the interacting case, we prove the remarkable fact that (with
normal ordering) the QIVP always has a unique solution and describe how
this solution may be constructed. To obtain more specific results, we
consider the cases in which ``space'' consists of either 2 or 3 points. 
In the 2-point model, we find that the unique solution to the QIVP
satisfies the CCR/CAR's to the future of the nonchronal region (so the
QIVP is well posed) and that in consequence the resulting quantum theory
is {\em unitary}. This contrasts strongly with the corresponding path
integral result (generalising that of Politzer~\cite{Pol2}) in which the
evolution is found to be
nonunitary. In consequence, and because the path integral also employs
normal ordering, we conclude that the self-consistent path integral
evolution does not generally correspond to a solution of the equation of
motion. Given the different starting points of the two approaches this
is not entirely surprising. 

In the 3-point model, we show that the QIVP is ill posed for both Bose
and Fermi statistics because the evolved data does not satisfy the
CCR/CAR's to the future of the dischronal region. The corresponding
quantum theory is therefore not unitary. We then
discuss the nature of this evolution in the fermionic case in order to
determine whether or not it can be described by means of a
superscattering operator. To do this
it is necessary to translate our results from the Heisenberg picture to
the Schr\"{o}dinger picture. Although there is no unique translation
prescription (as a consequence of the violation of the CAR's), we are 
nonetheless able to show that {\em no} Schr\"{o}dinger picture evolution
consistent with the QIVP solution can factorise into the product of an
operator and its adjoint, lending support to one element of Hawking's
position~\cite{Hawk}. However, it also transpires that no such
Schr\"{o}dinger picture evolution can be described by a superscattering
matrix as it must either {\em increase} the trace of density matrices or
map them to non-positive operators. In this sense, the loss of unitarity
in our model is much more radical than envisaged by Hawking. 

We also study the classical limit of our quantum theory. When this
limit exists uniformly one recovers exactly one solution to the
classical equation of motion. One might imagine that this
limit would fail when the classical theory is nonunique; however, this
is not the case. It appears that there are bands of ``coupling
strength'' for which the limit does exist even when there are many
classical solutions. We present numerical evidence to exemplify this
behaviour including an example in which the quantum theory selects
precisely one out of 25 classical solutions. We believe the convergence
bands continue to appear as the coupling strength is increased
indefinitely. Within the convergence bands, our quantum theory resolves
the classical nonuniqueness; for other values of the coupling strength,
it is arguable that {\em no} classical solution is physically relevant.

In addition, we consider the effect of altering the operator ordering
used and find that the solutions to the QIVP can become nonunique for
large quantum numbers for non-normal operator ordering. We study a
1-parameter family of operator orderings for the 3-point model and show
that the resulting quantum theories are all nonunitary. 

The paper is structured as follows. We first describe our class of
chronology violating models in Section~\ref{sect:prelim} and then
study the classical initial  
value problem for both linear and nonlinear fields in
Sect.~\ref{sect:class}. Next, in Sect.~\ref{sect:QwoC} we discuss the
quantum initial value problem for our models in the absence of CTC's and   
demonstrate its equivalence with canonical quantisation. This serves to
fix our notation and definitions for Section~\ref{sect:canon} in which
we uniquely solve the QIVP with CTC's present, and discuss the 2- and
3-point models, showing that the CCR/CAR's are violated in the 3-point
case. This nonunitary evolution is investigated in Sect.~\ref{sect:dnue}
and is shown not to be described by a superscattering operator.
Section~\ref{sect:classlim} treats the classical limit, whilst
Sect.~\ref{sect:fact} contains a brief discussion of the effect of
operator ordering on our results. In Sect.~\ref{sect:scpi}, we 
review the self-consistent path integral formalism, extending and in one  
instance correcting the treatment given by Politzer~\cite{Pol2}. We use
this formalism to compute the general (unitary) evolution 
for the free models, obtaining agreement with the QIVP. For
the 2- and 3-point interacting models we show that the QIVP and path
integral differ. We comment on this and other issues in
the Conclusion (Sect.~\ref{sect:concl}). There are three Appendices.
Appendix~\ref{sect:FQPI} reproduces our treatment of the free classical
evolution using the methods of Goldwirth {\em et al.}~\cite{GPPT},
whilst Appendix~\ref{sect:PTF} gives a derivation of the quantum
evolution of the free models using the formalism of
Politzer~\cite{Pol2}, rather than the more direct method employed in the 
text. In Appendix~\ref{sect:detail}, we present the details of a
calculation which shows that the CCR/CAR's are violated in the
3-point interacting model.

\section{A Class of Chronology Violating Models}
\label{sect:prelim}

In this section, we describe a class of nonlinear differential equations
on ``spacetimes'' in which ``space'' consists of finitely many discrete
points. By making identifications in these spacetimes, we
introduce CTC's and obtain spacetime models generalising that studied
by Politzer~\cite{Pol2}. These identifications are implemented in the
field theory by imposing certain boundary conditions which place
constraints on the theory.

Let ${\cal S}$ be a finite\footnote{ Many of our results for free fields 
generalise easily to the case in which ${\cal S}$ is a manifold.}
collection of points ${\cal S} = \{z_\alpha\mid \alpha =1,\ldots,s\}$
for some $s\ge 2$, 
and define spacetime to be the Cartesian product ${\cal S}\times {\Bbb
R}$. We define ${\frak H}$ to be the Hilbert space of complex-valued
functions on ${\cal S}$ with inner product $\langle f\mid g\rangle 
=\sum_{z\in {\cal S}} \overline{f(z)}g(z)$. This space has vectors
$v_\alpha$ as an orthonormal basis, where we define $v_\alpha(z_\beta)=
\delta_{\alpha\beta}$. With respect to this basis, we may write 
functions in ${\frak H}$ as $s$-dimensional complex vectors, so that
$\langle f\mid g\rangle = f^\dagger g= \overline{f_{\alpha}} g_\alpha$, 
where we sum over the repeated index.

We will study model field theories derived from Lagrangians of form
\begin{equation}
{\cal L} = \frac{i}{2}(\psi^\dagger\dot{\psi} - \dot{\psi^\dagger}
\psi) - \psi^\dagger W\psi -\frac{\lambda}{2}(\psi^\dagger\psi)^2,
\label{eq:Lag}
\end{equation}
where $\psi(t)\in {\frak H}$, $W$ is a self-adjoint positive operator 
on ${\frak H}$ and $\lambda\in{\Bbb R}^+$.  
The corresponding field equation is
\begin{equation}
\dot{\psi} = -i W\psi -i\lambda (\psi^\dagger\psi)\psi,
\label{eq:classf}
\end{equation}
which conserves the quantity
$\psi^\dagger\psi$, and therefore reduces to the {\em linear} equation
\begin{equation}
\dot{\psi} = -i W\psi - i\lambda\psi^\dagger(0)\psi(0)\psi,
\end{equation}
once the initial data $\psi(0)$ is specified. Thus the unique solution
to Eq.~(\ref{eq:classf}) with this data is
\begin{equation}
\psi(t) = e^{-i\lambda t \psi^\dagger(0)\psi(0)} e^{-i W t}\psi(0).
\label{eq:soln1}
\end{equation}

The configuration space variables $\psi_\alpha$ have conjugate momenta
$i\psi_\alpha^\dagger$,\footnote{Na\"{\i}vely, one might expect
the momenta to be $\frac{1}{2}i\psi^\dagger_\alpha$. However, the
Lagrangian~(\ref{eq:Lag}) is a second class constrained system and the
correct momenta may be obtained using Dirac brackets~\cite{Dirac}.} and 
Eq.~(\ref{eq:classf}) may be written in the Hamiltonian form 
\begin{equation}
\dot{\psi}_\alpha = \frac{\partial h}{\partial (i\psi^\dagger_\alpha)},
\end{equation}
with Hamiltonian
\begin{equation}
h(\psi,i\psi^\dagger) = \psi^\dagger_\alpha W_{\alpha\beta} \psi_{\beta} 
+ \frac{\lambda}{2}
\psi^\dagger_\alpha\psi^\dagger_\beta\psi_\beta\psi_\alpha .
\label{eq:Ham1}
\end{equation}

To introduce CTC's we partition ${\cal S}$ into two subsets
${\cal S}_1$ and ${\cal S}_2$ containing $s_1$ and $s_2$ elements
respectively, with $s_1+s_2=s$ and $s_2\le s_1$, and make pointwise 
identifications of ${\cal S}_2\times\{T^+\}$ with
${\cal S}_2\times\{0^-\}$ and ${\cal S}_2\times\{T^-\}$ with 
${\cal S}_2\times\{0^+\}$ for some $T>0$. This idealises wormholes
linking the lower surface of ${\cal S}_2$ at $t=0$ with the upper
surface of ${\cal S}_2$ at $t=T$, and the upper surface
of ${\cal S}_2$ at $t=0$ with the lower surface of ${\cal S}_2$ at
$t=T$. Note that $0^-$ and $0^+$ (and correspondingly $T^-$ and $T^+$) 
are regarded as distinct topological points for this purpose.

The partition of ${\cal S}$ induces a partition of the basis
vectors $v_\alpha$ into the sets $e_1,\ldots,e_{s_1}$ and $f_1,\ldots,f_{s_2}$
whose respective spans are denoted ${\frak H}_1$ and ${\frak H}_2$.
Clearly, we have $\dim{\frak H}_2\le \dim{\frak H}_1$. 
We will also write the projection of $\psi\in{\frak H}$ onto
${\frak H}_i$ ($i=1,2$) as $\psi_i$. In the classical field theory, 
the identifications are implemented by the imposition of the boundary
conditions 
\begin{equation}
\psi_2(T^-) = A\psi_2(0^+)\quad{\rm and}\quad\psi_2(T^+)=B\psi_2(0^-),
\label{eq:CTCbcs}
\end{equation}
where $A, B$ are unitary maps of ${\frak H}_2$ to itself, corresponding 
to the evolution through the wormholes 
(cf. Goldwirth {\em et al.}~\cite{GPPT}\footnote{In the notation 
of~\cite{GPPT}, our $A$ is equal to $(W^\downarrow)^{-1}$, and $B$ is
equal to $W^\uparrow$.}). Politzer~\cite{Pol2} takes $A=B=\openone$.
The r\^{o}le of these boundary conditions is simply to ensure 
that the evolution round the wormholes is consistent. We require
$\psi_1(t)$ to be everywhere continuous, although $\dot{\psi}_1(t)$
may be discontinuous at $t=0,T$. Thus~(\ref{eq:classf}) is suspended
at these points. 

Except in the special cases in which $\cal S$ consists either of 2
points or 3 points arranged in a ring, the interaction term
in~(\ref{eq:Lag}) is not a nearest neighbour interaction and is
therefore rather unsatisfactory as a model field theory. We will
therefore restrict our discussion of specific interacting models to
these cases. A more realistic field theory will be discussed
elsewhere~\cite{FWii}. 

\section{The Classical Initial Value Problem}

\label{sect:class}

In this section, we examine the behaviour of the classical field
equation~(\ref{eq:classf}) subject to the CTC boundary 
conditions~(\ref{eq:CTCbcs}). For a generic class of $W$ and $T$, we
show that the free field initial value problem is well posed
for data specified before the dischronal region. We then examine the
nonlinear theory and show that (generically) solutions exist for all
initial data specified before the dischronal region; moreover, this
solution is unique in the case of ``weak
coupling'', but nonunique for ``strong coupling''. 

To define the class of generic $W$ and $T$, we
decompose the operator $e^{-iWT}$ (which implements the free evolution
between $t=0^+$ and $t=T^-$) in the block form
\begin{equation}
e^{-iWT} = \left(\begin{array}{cc} P & Q \\ R & S \end{array}\right),
\label{eq:block}
\end{equation}
with respect to the decomposition ${\frak H}= {\frak H}_1\oplus {\frak
H}_2$. The {\em generic case} is defined to be the case in which the
norm\footnote{If $V_1$ and $V_2$ are complete normed linear
spaces (Banach spaces) with norms $\|\cdot\|_1$ and $\|\cdot\|_2$ and
$A:V_1\to V_2$ is a linear mapping, then the norm $\|A\|$ of $A$ is
defined by $\|A\|=\sup_{f\in V_1\backslash\{0\}} \|A f\|_2/\|f\|_1$.}
$\|S\|$ of $S$ is {\em strictly} less than unity. (Note that 
$\|S\|\le 1$ because $e^{-iWT}$ is unitary.) Equivalently, we require
that $Q$ should be an injection from ${\frak H}_2$ into ${\frak H}_1$ so 
that $Q$ has a left inverse $K$, (i.e., such that $KQ=\openone|_{{\frak
H}_2}$) which is uniquely specified if we require it to annihilate the
orthogonal complement of ${\rm Im}\, Q$. In this case we have
$\|K\|=\|Q\|^{-1}$. This requirement is the reason
for our restriction that $\dim {\frak H}_2\le
\dim{\frak H}_1$: otherwise, $Q$ would necessarily have nontrivial
kernel. The generic case corresponds to the situation expected in
physically realistic field theories in which wavepackets spread out so
that some proportion of any wave emerging from the wormhole at $t=0^+$
manages to avoid reentering it at $t=T^-$. 

\subsection{Free Case}
\label{sect:freec} 

We now show that, in the generic case with $\lambda=0$, the equation of
motion~(\ref{eq:classf}) with boundary conditions~(\ref{eq:CTCbcs})
constitutes a well posed initial value problem for arbitrary data 
$\psi\in{\frak H}$ specified at $t=0^-$ (and therefore for any $t<0$).
In fact, we will only need the weaker condition that $A-S$ be
invertible on ${\frak H}_2$.

The evolution between $t=0^+$ and $t=T^-$ is given simply by the
operator $e^{-iWT}$; accordingly, given data at $t=0^-$, the problem
reduces to the study of the evolution between $t=0^-$ and $t=0^+$.
Because $\psi_1(t)$ is required to be continuous at $t=0$, it remains to 
determine $\psi_2(0^+)$ in terms of $\psi(0^-)$. The only constraint on
$\psi_2(0^+)$ is that the CTC boundary conditions be satisfied, i.e.,
that  $\psi_2(T^-)=A \psi_2(0^+)$. From Eq.~(\ref{eq:block}) we have
$\psi_2(T^-)= R\psi_1(0)+S\psi_2(0^+)$ so, provided $A-S$ is invertible,  
$\psi_2(0^+)$ is uniquely specified as
\begin{equation}  
\psi_2(0^+) = (A - S)^{-1} R \psi_1(0) .
\label{eq:consreq}
\end{equation}

For $0<t<T$, the solution is thus
\begin{equation}
\psi(t) = e^{-iWt}\left(\begin{array}{c} \psi_1(0) \\
(A-S)^{-1}R\psi_1(0) 
\end{array}\right),
\end{equation}
and in particular, we obtain
\begin{equation}
\psi_1(T) =  M\psi_1(0) ,
\end{equation}
where the matrix $M$ is
\begin{equation}
M = P + Q (A - S)^{-1} R .
\end{equation}
The matrix $M$ is unitary, as we now show. Let $\psi\in{\frak H}_1$ and
use the unitarity of $e^{-iWT}$ to compute
\widetext
\begin{eqnarray}
\|\psi\|^2+\|(A-S)^{-1}R\psi\|^2 &=& \left\| e^{-iWT}\left(
\begin{array}{c} \psi \\ (A-S)^{-1}R\psi \end{array} \right) \right\|^2
\nonumber \\
&=& \| (P+Q(A-S)^{-1}R) \psi\|^2 + \|(\openone + S(A-S)^{-1})R\psi\|^2
\nonumber \\
&=& \| M\psi\|^2 + \|A(A-S)^{-1}R\psi\|^2 .
\end{eqnarray}
\narrowtext
By unitarity of $A$, we now have $\|M\psi\|^2=\|\psi\|^2$ and conclude
that $M$ is unitary. 

Thus we have shown that there is a unique classical solution for each
choice of initial data at $t=0^-$ and that the full classical evolution
from $t=0^-$ to $t=T^+$ is
\begin{equation}
\psi(T^+) = \left(\begin{array}{cc} M & 0 \\ 
0 & B \end{array}\right) \psi(0^-).
\label{eq:classbc}
\end{equation}
Moreover, the solution is clearly continuous in the initial data,
so we conclude that this initial value problem is well posed for data
specified in the past of the CTC region on surfaces of constant $t$.

The situation is different for data specified between $t=0^+$ and
$t=T^-$. Here, the initial value problem is well posed only for a
subclass of data satisfying certain consistency requirements. For
example, data specified at $t=0^+$ must obey Eq.~(\ref{eq:consreq}).
This phenomenon has been noted before in various
situations~\cite{FMNEKTY,GPPT,Yurt}; it arises because the CTC's
introduce 
constraints on the dynamics and has important implications for the
quantum theory. Note that one may nonetheless specify the data at any
given point freely: it is always possible to choose the remaining
initial data so as to satisfy the consistency requirements. Thus our
system has a ``benignity'' property analogous to those discussed
in~\cite{FMNEKTY,Yurt}. Related to this phenomenon is the fact that the
classical evolution is
{\em nonunitary} between $t=0^-$ and $0^+$ and between $t=T^-$ and
$T^+$. To see this, take any initial data with $\psi_1(0)=0$: at
$t=0^-$, the initial data has norm $\|\psi_2(0^-)\|$; for $0<t<T$
the solution vanishes identically; and finally, at $t=T^+$, the solution
again has norm $\|\psi_2(0^-)\|$.

Finally, it is instructive to see how this classical evolution may be 
derived using the path integral methods of Goldwirth 
{\em et al.}~\cite{GPPT}. This is described in Appendix~\ref{sect:FQPI}.

\subsection{Interacting Case}
\label{sect:mod1cl}

We now consider the full interacting classical field theory in the
generic case. We will show: (i) there exists at least one solution
for arbitrary initial data; (ii) there is a weak coupling regime in
which there is a {\em unique} solution; and (iii) there is a strong
coupling regime in which there exist many distinct solutions for each
choice of initial data. 

In the absence of CTC boundary conditions, the solution is given
by~(\ref{eq:soln1}). We write $a=\psi_1(0)$ and $b = \psi_2(0^+)$
and implement the CTC boundary conditions by requiring $b$ to satisfy 
\begin{equation}
A b = e^{-i\lambda T (a^\dagger a + b^\dagger b)}(Ra+Sb),
\label{eq:bcon}
\end{equation}
for given $a$.

To study the solutions to this equation, we first note that it implies
$\|b\| = \|Ra+Sb\|$ and hence, by the unitarity of $e^{-iWT}$, that
$\|a\| = \|Pa+Qb\|$. In the generic case (in which $Q$ has left inverse 
$K$) any solution $b$ must therefore take the form
\begin{equation}
b = K(U - P) a ,
\label{eq:bfrm}
\end{equation}
for some unitary $U$ on ${\frak H}_1$. Substituting back into
Eq.~(\ref{eq:bcon}), and rearranging, we find that $b$
solves~(\ref{eq:bcon}) if and only if
\begin{equation}
KUa = Kf(U) a ,
\end{equation}
where $f(U)=P+Q(Ae^{i\eta(U)}-S)^{-1}R$ and 
\begin{equation}
\eta(U) = \lambda T a^\dagger\left\{\openone +(U-P)^\dagger
K^{\dagger}K (U-P)\right\}a.
\end{equation}
Because $\eta(U)$ is real-valued, $f(U)$ is a unitary operator on
${\frak H}_2$. 

Clearly, any solution to the fixed
point equation $U=f(U)$ necessarily yields a solution to
Eq.~(\ref{eq:bcon}); moreover, any such $U$ must take the form
$U(z)=P+Q(zA-S)^{-1}R$ for some $z$ on the unit circle. (Note that this 
expression is always well-defined because $\|S\|<1$.) Thus the
problem of existence reduces to finding fixed points of the equation
$z=e^{i\eta(U(z))}$ on the unit circle. Now 
\begin{equation}
\eta(U(z)) = \lambda T \left( \|a\|^2+\|(zA-S)^{-1}R a\|^2 \right) ,
\end{equation}
which is a continuous single-valued function from the unit circle to
the real line; thus $e^{i\eta(U(z))}$ is a mapping of the unit circle to    
itself with vanishing Brouwer degree (see, e.g., Ch.~1 of~\cite{Deim}). 
Accordingly, for each choice of initial
data $a\in{\frak H}_1$ there exists at least one fixed point of $g$ and
thus at least one solution to Eq.~(\ref{eq:bcon}), so we have proved the 
claim~(i) above.

To establish claim~(ii), we write the right hand side of
Eq.~(\ref{eq:bcon}) as $A g(b)$ where $g:{\frak H}_2\to{\frak H}_2$ 
and consider the fixed point problem $b=g(b)$ on the ball 
${\cal B}=\{b\in{\frak H}_2\mid\|b\|\le r_0\|a\|\}$,
where $r_0=\|Q\|^{-1}(1+\|P\|)$. This ball contains all solutions to
Eq.~(\ref{eq:bcon}) as a consequence of Eq.~(\ref{eq:bfrm}). 
For any $b_1,b_2$ in $\cal B$, we have
\begin{eqnarray}
\|g(b_1)-g(b_2)\| &=& \left\| (1-e^{i\lambda T
(\|b_1\|^2-\|b_2\|^2)}) (Ra + Sb_1) \right. \nonumber \\
&& \left. + e^{i\lambda T
(\|b_1\|^2-\|b_2\|^2)}S(b_1- b_2)\right\| \nonumber \\
&\le & \left(  c_0\lambda T\|a\|^2 +\|S\| \right) \|b_1-b_2\|,
\end{eqnarray}
in which we have used the elementary estimates $|1-e^{i\alpha}|\le
|\alpha|$ and $| \|b_1\|-\|b_2\| |\le \|b_1-b_2\|$, and
$c_0=2r_0 (\|R\|+\|S\|r_0)$ is a positive real constant depending only
on $P,Q,R$ and $S$. In the generic case, for $\lambda T\|a\|^2 <
c_0^{-1}(1-\|S\|)$  (i.e.,
weak coupling), $g|_{\cal B}$ is a strict contraction (which need not
map ${\cal B}$ to itself) and standard contraction mapping arguments now 
imply that there can be at most one fixed point in ${\cal B}$. Putting
this together with~(i) and using the fact that all fixed points of $g$
must lie in ${\cal B}$, we have proved~(ii). 

Finally, to prove~(iii) we note that if $\lambda T \|Ra\|^2\gg 1$ (i.e.,
strong coupling) then the fixed point problem $z=e^{i\eta(U(z))}$
described above has many solutions on the unit circle; moreover, because
$Ra\not=0$, these solutions must correspond to distinct values of $b$ and
hence of $\psi_1(T)$. 

\begin{figure}
\center{\leavevmode\epsfxsize=6truein\epsfbox{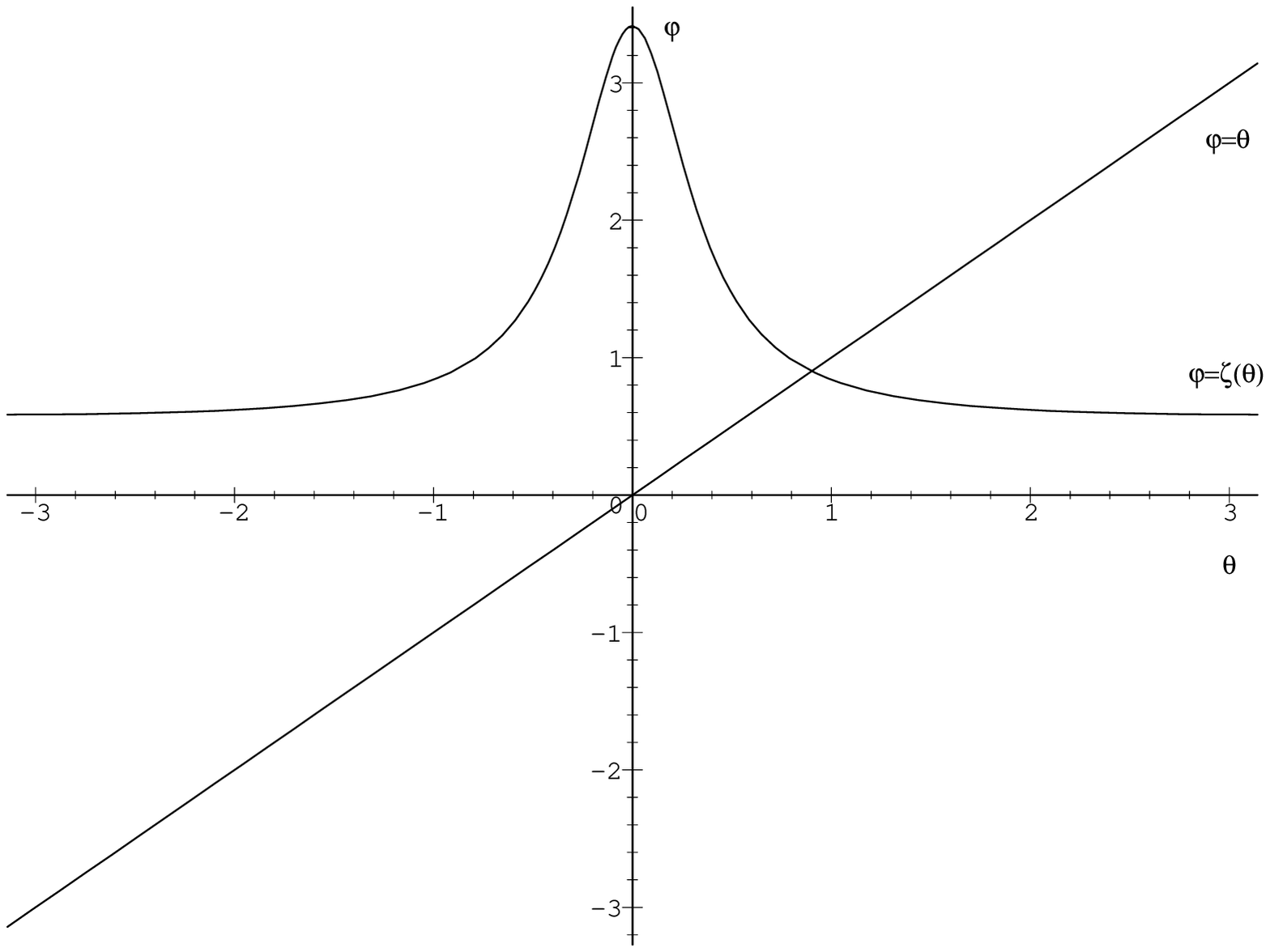}}
\caption{Graphical solution of Eq.~(\protect\ref{eq:grsol}) for 
$\mu=0.5$ showing that there is a unique solution of the classical
solution for this coupling strength.}
\label{fig1}
\end{figure}

\begin{figure}
\center{\leavevmode\epsfxsize=6truein\epsfbox{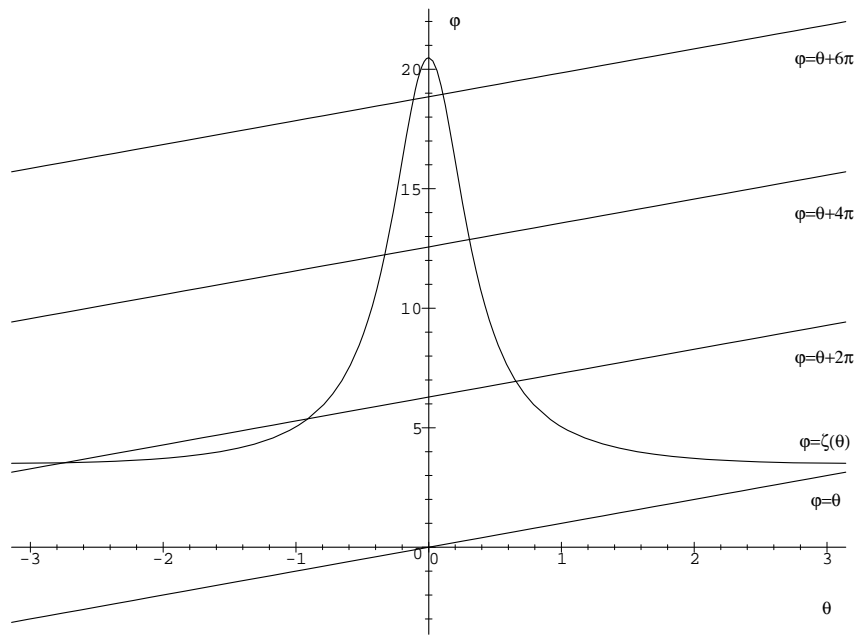}}
\caption{Graphical solution of Eq.~(\protect\ref{eq:grsol}) for 
$\mu=3.0$ showing that there are $7$ solutions for
this coupling strength.}
\label{fig2}
\end{figure}

In Figs.~\ref{fig1} and~\ref{fig2}, we explicitly show how nonuniqueness
arises in a model with two spatial points and $P=-Q=R=S=1/\sqrt{2}$ with
$A=\openone$. 
For this model the classical solutions are in one-to-one correspondence
with the solutions of the fixed point equation $z=e^{i\eta(U(z))}$ on the
unit circle, because $K$ and $U(z)$ are scalars. Writing
$z=e^{i\theta}$, this becomes
\begin{equation}
\zeta(\theta)= \theta + 2k\pi ,
\label{eq:grsol}
\end{equation}
for $k\in{\Bbb Z}$ where $\zeta(\theta)= \mu (1+|\sqrt{2}
e^{i\theta}-1|^{-2})$ and we have written $\mu=\lambda T|a|^2$ for
the ``coupling strength''. This equation may be solved graphically by
plotting each side separately for $-\pi<\theta\le\pi$ looking for
intersections. Fig.~\ref{fig1} shows the appropriate plots for
$\mu=0.5$, from which it is clear that there is a unique solution,
whilst Fig.~\ref{fig2} corresponds to the case $\mu=3.0$ where there
are $7$ solutions.

\subsection{Preservation of the Symplectic Structure}

Except at $t=0, T$, the classical dynamics is generated by the
Hamiltonian $h$, and therefore preserves the symplectic structure on
phase space in the initial and final chronal regions. Owing to the 
CTC boundary conditions, it is not clear that the evolution from initial
to final chronal regions also preserves the symplectic structure. Here,
we express the classical evolution in phase space language
and prove that the evolution from initial to final chronal regions is
implemented by a symplectic transformation. 

The classical phase space is $\Gamma={\Bbb C}^s$ with symplectic
structure given by the 2-form $\Omega=-i d\psi^\dagger\wedge 
d\psi=-id\overline{\psi}_\alpha \wedge d\psi_\alpha$.
In the usual way, functions on $\Gamma$ are regarded as functions of
independent variables $\psi$ and $\psi^\dagger$. A symplectic
transformation $\xi$ is a diffeomorphism of $\Gamma$ which preserves
$\Omega$, i.e., $\xi^*\Omega=\Omega$, where
$\xi^*\Omega$ is the pull-back of $\Omega$ by $\xi$. This is 
equivalent (see e.g., \S 40 in~\cite{Arnold}) to the Dirac bracket
relation 
\begin{equation}
\left\{ f\circ\xi, g\circ\xi\right\}_{D,x} =
\left\{ f,g\right\}_{D,\xi(x)} ,
\end{equation}
where the Dirac bracket of two functions on $\Gamma$ is
defined by
\begin{equation}
\left\{f,g\right\}_{D,x} =\left.\left(
\frac{\partial f}{\partial \psi_\gamma}  
\frac{\partial g}{\partial i\overline{\psi}_\gamma} - 
\frac{\partial g}{\partial \psi_\gamma}  
\frac{\partial f}{\partial i\overline{\psi}_\gamma}\right)\right|_{x} .
\end{equation} 
 
Corresponding to the decomposition $\psi=(\psi_1,\psi_2)$, we have
$\Gamma=\Gamma_1\times\Gamma_2$, and associated natural projections
$\pi_k:\Gamma\to\Gamma_k$. Then $\Omega$ can be expressed as
\begin{equation}
\Omega = \pi_{1}^*\Omega_1 + \pi_{2}^*\Omega_2,
\end{equation}
where, for $k=1,2$, $\Omega_k=-id\psi_k^\dagger\wedge d\psi_k$ is the 
symplectic form on $\Gamma_k$. Any unitary matrix $U$ on ${\frak
H}_2$ defines a corresponding natural symplectic transformation of
$\Gamma_2$, which we denote $\chi_U$. In addition, for $t\in{\Bbb R}$,
$\tau_t=\exp tIdh$ is the evolution generated by the Hamiltonian
$h$, where $I$ is the canonical isomorphism between 1-forms and vector  
fields on $\Gamma$ specified by $\Omega(I\omega, \cdot) =
\omega(\cdot)$. We have $\tau_t^*\Omega=\Omega$ for all $t$.

With these definitions, the diffeomorphism $\eta$ implementing evolution 
from $t=0^-$ to $t=T^+$ is $\eta = (\kappa,\chi_B)$, where
$B$ is the unitary matrix appearing in the CTC boundary conditions and
$\kappa$ is a mapping from an open set $U \subset \Gamma_1$ into
$\Gamma_1$ defined as follows. First, we define a
differentiable map $\sigma:U\to \Gamma$ as a solution to the equations 
\begin{equation}
\pi_1\circ\sigma  =  {\rm id}_1,\ \ \  
\pi_2\circ\tau_T\circ\sigma = \chi_A\circ\pi_2\circ\sigma, 
\label{eq:rel1} 
\end{equation}
which express the classical consistency requirement Eq.~(\ref{eq:bcon}).
Then $\kappa$ is defined by
\begin{equation}
\kappa   =  \pi_1\circ\tau_T\circ\sigma.  \label{eq:rel2}
\end{equation}
In general, there will be many possible choices for $\kappa$ reflecting
the nonuniqueness of the classical evolution. For any such choice,
the relation $\kappa^*\Omega_1 = \Omega_1$ can be proved 
using the composition rule
$(f\circ g)^* = g^* f^*$ as
\begin{eqnarray}
\kappa^*\Omega_1 & = & \sigma^*\tau_T^*\pi_1^*\Omega_1 \nonumber \\
   & = & \sigma^*\tau_T^*(\Omega - \pi_{2}^{*}\Omega_2) \nonumber \\
   & = & \sigma^*\Omega - \sigma^*\tau_T^*\pi_2^*\Omega_2 \nonumber \\
   & = & \sigma^*(\pi_1^*\Omega_1 + \pi_2^*\Omega_2)
   - \sigma^*\pi_2^*\chi_B^*\Omega_2 \nonumber \\
   & = & \Omega_1.
\end{eqnarray}
Thus, because $\chi_B^*\Omega_2=\Omega_2$, we conclude that
$\eta=(\kappa,\chi_B)$ preserves $\Omega$. 

In terms of Dirac brackets, writing
$\psi(T^+)=\eta(\psi,i\psi^\dagger)$, we have proved 
\begin{equation}
\left\{ \psi_\alpha(T^+), \psi_\beta(T^+)\right\}_D =0,
\end{equation}
and
\begin{equation}
\left\{ \psi_\alpha(T^+), \overline{\psi}_\beta(T^+) \right\}_D =
-i\delta_{\alpha\beta} .
\end{equation}

Note that the evolution between $t=0^-$ and $t=0^+$ (and similarly
between $t=T^-$ and $t=T^+$) is {\em not} symplectic in general.

\section{The Quantum Initial Value Problem In the Absence of CTC's}
\label{sect:QwoC}

In order to prepare for our discussion of chronology violating models,
it is useful to show how a study of the QIVP reproduces the results of
canonical quantisation for Eq.~(\ref{eq:classf}) in the absence of
CTC's. We first discuss the case of Fermi statistics to avoid
the operator domain technicalities of the bosonic case. 

The canonical approach starts by identifying the classical
canonical coordinates $\psi_\alpha$ and $i\psi_\alpha^\dagger$ and the
classical Hamiltonian
$h(\psi_\alpha,i\psi_\alpha^\dagger)$ defined in Eq.~(\ref{eq:Ham1}). 
A Hilbert space ${\frak F}$ is then constructed on which bounded
operators $\Psi_1,\ldots,\Psi_s$ represent the CAR's
for $s$ degrees of freedom -- that is, $\{\Psi_\alpha,\Psi_\beta\} = 0$
and $\{\Psi_\alpha,\Psi_\beta^\dagger\} = \delta_{\alpha\beta}$ for all
$\alpha,\beta$. The quantised (normal ordered) Hamiltonian $H$ is
defined as a (bounded)
self-adjoint operator on $\frak F$ by substituting $\Psi_\alpha$ for
$\psi_\alpha$ in the RHS of Eq.~(\ref{eq:Ham1}) using its literal
ordering. The quantum evolution generated by $H$ evolves a general
operator $A$ from time $0$ to $t$ by
\begin{equation}
A(t) = e^{iHt} A e^{-iHt} ,
\label{eq:Heisev}
\end{equation}
and the evolved operator therefore satisfies the Heisenberg equation of  
motion 
\begin{equation}
\dot{A}(t) = i[H,A(t)] .
\end{equation}
Thus, by virtue of the CAR's, 
$\Psi_\alpha(t)=e^{iHt}\Psi_\alpha e^{-iH t}$ solves the original
equation of motion~(\ref{eq:classf}) 
as an operator differential equation with initial data
$\Psi_\alpha(0)= \Psi_\alpha$. Moreover, the CAR's are necessarily
preserved by this evolution.

It is possible to reproduce these results from a slightly different
angle, namely by treating Eq.~(\ref{eq:classf}) as an operator
differential equation and considering the {\em quantum initial value
problem} (QIVP). Given initial data $\Psi_\alpha$ representing the
CAR's, we say that the QIVP is well posed if there exists a unique
operator-valued solution $\Psi_\alpha(t)$ to Eq.~(\ref{eq:classf}) with
$\Psi_\alpha(0)=\Psi_\alpha$ and the evolution preserves the CAR's. To
show that this is indeed the case, we note
that for {\em arbitrary} initial data given as bounded operators on
${\frak F}$, Eq.~(\ref{eq:classf}) has the unique solution
\begin{equation}
\Psi_\alpha(t) =
e^{-i\lambda t \Psi_\gamma^\dagger\Psi_\gamma}
\left( e^{-iWt}\right)_{\alpha\beta} \Psi_\beta  .
\end{equation}
The proof of uniqueness closely parallels the analogous argument for the 
classical differential equation. One may check that this evolution
preserves the CAR's either by explicit computation or by noting that 
the above solution must agree (by uniqueness) with that obtained from
the canonical approach. Thus the QIVP for Eq.~(\ref{eq:classf}) is well
posed in the fermionic case. 

Of course, it is not usually advantageous to consider the QIVP directly
because it is rare that the equation of motion may be solved in closed
form for general operator-valued initial data. However, for the
chronology violating models considered in this paper, it will not always
be possible
to assume that the initial data is a representation of the canonical
(anti)commutation relations and therefore the canonical method is no
longer guaranteed to yield solutions to the equation of motion
Eq.~(\ref{eq:classf}). In these situations, we must therefore employ
the more general setting of the QIVP.   

In the bosonic case, we encounter unbounded operators and therefore must 
proceed more carefully. We now describe the technicalities required in
order to generalise the foregoing to this case.\footnote{For
completeness, we give the following definitions (see~\cite{RSi}). An
operator $A$ with dense domain $D(A)$ in Hilbert space $\frak F$ is {\em
closed} if its graph (i.e., the set of pairs $\langle f,Af\rangle$ as
$f$ runs through $D(A)$) is a closed subspace of ${\frak F}\times {\frak
F}$ in the product topology; $A$ is closable if the closure of its graph
is itself the
graph of an operator, called the {\em closure} of $A$. An algebraic
subspace ${\cal D}$ of ${\frak F}$ contained in $D(A)$ is a {\em core}
for a closed operator $A$ if $A$ is the closure of its restriction to
$\cal D$. A densely defined operator $A$ is {\em essentially
self-adjoint} if its closure is self-adjoint. ${\cal D}\subset
D(A)$ is {\em invariant} under $A$ if $Af\in{\cal D}$ for all $f\in{\cal
D}$. Finally, an operator-valued
function $A(t)$ is {\em strongly differentiable} with respect to $t$ on
$\cal D$ with derivative $B(t)$ if $\cal D$ is contained in $D(B(t))$
and $D(A(\tau))$ for all $\tau$ in some neighbourhood of $t$ and
$\|(\epsilon^{-1}(A(t+\epsilon)-A(t))-B(t))f\| \rightarrow 0$ as
$\epsilon\rightarrow 0$ for all $f\in{\cal D}$.}

{\noindent\bf Definition}\,{\em Let $\cal D$ be dense in Hilbert space
$\frak F$, and let $\Psi_1(t),\ldots,\Psi_s(t)$ be closed
operator-valued functions on ${\Bbb R}$ such that $\cal D$ is a core for 
each $\Psi_\alpha(t)$ and is invariant under the $\Psi_\alpha(t)$ and
$\Psi_\alpha^\dagger(t)$. Then the $\Psi_\alpha(t)$ are said to be a
solution to Eq.~(\ref{eq:classf}) on $\cal D$ if each
$\Psi_\alpha(t)$ is strongly differentiable with respect to $t$ on $\cal 
D$ with derivative $-iW_{\alpha\beta}\Psi_\beta(t) -i\lambda
\Psi_\gamma(t)^\dagger \Psi_\gamma(t)\Psi_\alpha(t)$.}

Note that this definition extends that used above
for the bounded case. 

{\noindent\bf Definition}\,{\em The closed operators
$\Psi_1,\ldots,\Psi_s$ are said to represent the CCR's on
$\frak F$ if they have a common dense domain $\cal X$ invariant under
both the $\Psi_\alpha$ and the $\Psi_\alpha^\dagger$ with 
\begin{equation}
[\Psi_\alpha,\Psi_\beta]f = [\Psi_\alpha^\dagger,\Psi_\beta^\dagger]f =0
\end{equation}
and
\begin{equation}
[\Psi_\alpha,\Psi_\beta^\dagger]f =
\delta_{\alpha\beta}f 
\end{equation}
for all $f\in{\cal X}$ and such that $\Psi_\alpha^\dagger\Psi_\alpha$
(summing over the repeated index) is decomposable on ${\cal X}$. 
That is, there exists a projection-valued measure $P_{\Omega}$ on 
$\Bbb R$ such that $\cal X$ contains ${\cal D}_0 = \bigcup_{\mu\ge 0} 
P_{[-\mu,\mu]}{\frak F}$ and $\Psi_\alpha^\dagger \Psi_\alpha f = 
\int_{\Bbb R} \mu dP_\mu f$ for all $f\in{\cal X}$.}

The reason for the technical requirement of decomposability is that it
guarantees~\cite{Rellich} that all such
representations of the CCR's are equivalent up to unitary equivalence
and multiplicity (i.e., the conclusion of von Neumann's theorem holds).

The canonical quantisation of Eq.~(\ref{eq:classf}) proceeds as follows.
Suppose that operators $\Psi_\alpha$ represent the CCR's on Hilbert
space $\frak F$ with dense invariant domain $\cal X$, and let ${\cal
D}_0\subset{\cal X}$ be defined as above. The quantum Hamiltonian may
be defined on ${\cal D}_0$ by substituting the operators $\Psi_\alpha$ into
the RHS of Eq.~(\ref{eq:Ham1}) to yield an essentially self-adjoint
operator whose closure is denoted by $H$. Moreover, ${\cal D}_0$ is easily
seen to be invariant under $e^{iHt}$ for $t\in{\Bbb R}$. Thus, the
evolved operators $\Psi_\alpha(t)$ defined by
\begin{equation}
\Psi_\alpha(t) = e^{iHt} \Psi_\alpha e^{-iHt}
\end{equation}
are strongly differentiable with respect to $t$ on ${\cal D}_0$
with derivative $ie^{iHt}[H,\Psi_\alpha]e^{-iHt}$ and the CCR's may
then be used (on ${\cal D}_0$) to conclude that the $\Psi_\alpha(t)$ solve
Eq.~(\ref{eq:classf}) on ${\cal D}_0$ in the sense defined above. 

As in the fermionic case, we may reproduce these results by studying the
QIVP. The situation for general initial data is summarised by the
following:

{\noindent\bf Proposition}\,{\em Let ${\frak F}$ be a Hilbert space and
${\cal D}\subset{\frak F}$ be dense. Suppose that $\Psi_\alpha$,
($\alpha = 1,\ldots,s$) are closed (possibly unbounded) operators on
$\frak F$ such that
\begin{enumerate}
\item[(i)] ${\cal D}$ is a core for each $\Psi_\alpha$ and a domain of
essential self-adjointness
for $\Psi_\gamma^\dagger\Psi_\gamma$
\item[(ii)] ${\cal D}$ is invariant under $\Psi_\alpha$,
$\Psi_\alpha^\dagger$ and 
$e^{-i\lambda t \Psi_\gamma^\dagger\Psi_\gamma}$ for all $t\in{\Bbb R}$.
\end{enumerate}
Then the operators $\Psi_\alpha(t)$ defined as the closure of 
$e^{-i\lambda t \Psi_\gamma^\dagger\Psi_\gamma}
\left( e^{-iWt}\right)_{\alpha\beta} \Psi_\beta$ on $\cal D$ constitute
the unique solution to Eq.~(\ref{eq:classf}) on $\cal D$ with initial
data $\Psi_\alpha$.} 

An immediate corollary of this is that if the $\Psi_\alpha$ represent
the CCR's on $\frak F$ and the domain ${\cal D}_0$ is defined as above, then
the QIVP for Eq.~(\ref{eq:classf}) is well posed on ${\cal D}_0$.

\section{The Quantum Initial Value Problem for Chronology Violating Models}
\label{sect:canon}

\subsection{General Formalism}

We now analyse the quantum initial value problem for
Eq.~(\ref{eq:classf}) in the presence of CTC's, beginning with the case
of the CAR's. Suppose that the operators $\Psi_\alpha$
($\alpha=1,\ldots,s$) provide a representation of the CAR's on Hilbert
space $\frak F$. We specify these operators as the initial data for the
QIVP at time $t=0^-$. Writing $\Psi_{1,i}$ and $\Psi_{2,j}$ to denote
those operators associated with ${\cal S}_1$ and ${\cal S}_2$
respectively, we therefore seek operators $\Psi_{2,j}(0^+)$ such that
the evolution between $t=0^+$ and $T^-$ obeys the consistency
requirement $\Psi_{2,i}(T^-)=A_{ij}\Psi_{2,j}(0^+)$. Denoting 
$\Psi_{1,i}=a_i$, $\Psi_{2,j}(0^+)=b_j$ we therefore require the $b_i$
to satisfy
\begin{equation}
A_{ij}b_j = e^{-i\lambda T(a_k^\dagger a_k +b_k^\dagger b_k)}\left(
R_{ij}a_j + S_{ij}b_j\right) .
\label{eq:b}
\end{equation}
Remarkably, and in contrast to the situation for the classical theory,
it turns out that this specifies the $b_i$ uniquely in the generic
case as we now show.

We first construct a solution to Eq.~(\ref{eq:b}) and then prove its
uniqueness. For $z\in {\Bbb C}$, let $N(z)_{ij}$ be the 
matrix-valued function of $z$ defined by $N(z)=(z A-S)^{-1}R$, which is  
analytic in an open neighbourhood of the unit circle in the generic
case. Then for any unitary operator $V$ on Hilbert space ${\frak K}$,
we may use the (Dunford) functional calculus (see e.g., pp. 556-577
of~\cite{DS}) to define
$N(V)_{ij}$ as a matrix of bounded operators on ${\frak K}$. Using this 
notation, Eq.~(\ref{eq:b}) may be rewritten in the form
\begin{equation}
b_{i} = N(e^{i\lambda T(a_k^\dagger a_k+b_k^\dagger b_k)})_{ij}a_j .
\label{eq:brew}
\end{equation}
Next, let ${\frak F}_r$ be the eigenspace of $a_i^\dagger a_i$ with
eigenvalue $r$ and decompose ${\frak F}=\bigoplus_r {\frak F}_r$. We
emphasise that $a_i^\dagger a_i$ is not the total particle number on
${\cal S}$ at $t=0^-$, but rather the particle number on ${\cal S}_1$.
Thus, for example, ${\frak F}_0$ is not 1-dimensional, but consists of
all states at $t=0^-$ with no ${\cal S}_1$-particles. 
We now define unitary operators $U_r$
on the ${\frak F}_r$ by the recurrence relation
\begin{equation}
U_{r+1} = \exp i\lambda T a^\dagger_i\left(\delta_{ik}+
N(U_r)^\dagger_{ij}N(U_r)_{jk}\right) a_k ,
\label{eq:recrl}
\end{equation}
with $U_0=\openone$. Denoting $U=\bigoplus_r U_r$, it is easy to see
that Eq.~(\ref{eq:b}) is solved by
\begin{equation}
b_i = N(U)_{ij}a_j ,
\label{eq:usol}
\end{equation}
by comparing with Eq.~(\ref{eq:brew}) and using the fact that each $a_i$
maps ${\frak F}_{r+1}$ to ${\frak F}_r$ and annihilates ${\frak F}_0$.

We now prove that~(\ref{eq:usol}) is the unique solution to
Eq.~(\ref{eq:b}). Suppose that $b_1,\ldots,b_{s_2}$ solve
Eq.~(\ref{eq:b}), and write $U=e^{i\lambda
T (a_k^\dagger a_k + b_k^\dagger b_k)}$. Because $N(U)_{ij}$ is a matrix
of bounded operators, Eq.~(\ref{eq:brew}) implies that the $b_i$
annihilate ${\frak F}_0$. Accordingly, $U$ leaves ${\frak F}_0$
invariant and $U|_{{\frak F}_0}=\openone$. Now suppose inductively that 
$U$ leaves ${\frak F}_r$
invariant for some $r\ge 0$. Provided that $r$ is not the largest
eigenvalue of $a_k^\dagger a_k$, Eq.~(\ref{eq:brew}) and its adjoint
imply that each $b_i$ maps ${\frak F}_{r+1}$ to ${\frak F}_r$ and each  
$b_i^\dagger$ maps ${\frak F}_r$
to ${\frak F}_{r+1}$. Accordingly $a_k^\dagger a_k+b_k^\dagger b_k$ and
thus $U$ leave ${\frak F}_{r+1}$ invariant. Hence by induction, we find
that each ${\frak F}_r$ is an invariant subspace for $U$, so we may
write $U=\bigoplus_r U_r$ with each $U_r$ unitary on ${\frak F}_r$. 
It is then easy to see that the $U_r$ must
satisfy the recurrence relation~(\ref{eq:recrl}) with $U_0=U|_{{\frak
F}_0}=\openone$. We have therefore completed the proof of uniqueness.

Finally, we note that this solution is representation independent in the
following sense. Suppose that $\Psi_\alpha$ form a Fock representation
of the CAR's, and let $b_i$ be the unique solution to Eq.~(\ref{eq:b})
on ${\frak F}$. By the Jordan--Wigner theorem (see e.g., \cite{Sim}),
an arbitrary representation $\Psi_\alpha'$ on ${\frak F}'$
takes the form $\Psi_\alpha'= U^{-1}(\Psi_\alpha\otimes \openone) U$,
where $U:{\frak F}'\rightarrow{\frak F}\otimes{\frak N}$ is unitary and
${\frak N}$ is an auxiliary Hilbert space. Then the unique solution to
the analogue of Eq.~(\ref{eq:b}) on ${\frak F}'$ is $b_i' =
U^{-1}(b_i\otimes \openone)U$. 

In the CCR case, certain domain questions must be addressed. We suppose
that the $\Psi_\alpha$ are a representation of the CCR's on ${\frak F}$ 
with common invariant domain $\cal X$ and define ${\cal D}_0\subset{\cal X}$
as in Section~\ref{sect:QwoC}. An important property of this domain is
that ${\frak F}_r\subset {\cal D}_0$ for all $r$, where ${\frak F}_r$ is
again defined as the
eigenspace of $a_i^\dagger a_i$ with eigenvalue $r$. Then it is easy to
see that the same construction as used in the CAR case yields a solution
to Eq.~(\ref{eq:b}) on ${\cal D}_0$; moreover, one may show that it is the
unique solution such that ${\cal D}_0$ is a core for each $b_i$, and is
independent of representation in the same sense as in the CAR case.

Once the unique solution to Eq.~(\ref{eq:b}) has been obtained (for
either CAR's or CCR's) we may substitute back to find 
\begin{equation}
a_i(T) = e^{-i\lambda T(a^\dagger_k a_k+b^\dagger_k b_k)}
\left(P_{ij} a_j + Q_{ij} b_j \right) ,
\end{equation}
and check to see whether or not this evolution preserves the CCR/CAR's
and is therefore unitary. We will analyse various cases of this problem
in the following subsections.

\subsection{Free Fields}

Here $\lambda = 0$ and Eq.~(\ref{eq:brew}) immediately yields
the unique solution
\begin{equation}
b_i = \left( A - S\right)^{-1}_{ij} R_{jk}a_k.
\end{equation}
Substituting, we find that the evolution is given by
\begin{eqnarray}
\Psi_{1,i}(T) &=& M_{ij}\Psi_{1,j}(0) \nonumber \\ 
\Psi_{2,i}(T^+) &=& B_{ij} \Psi_{2,j}(0^-),
\end{eqnarray}
where $M=P + Q( A- S)^{-1}R$ is unitary. Note that one obtains the same
result for both Bose and Fermi statistics. This evolution is easily seen
to preserve the CCR/CAR's; there is therefore a unitary $X$ on $\frak F$ 
such that
\begin{equation}
\Psi_\alpha(T^+) = X^\dagger \Psi_\alpha(0^-) X.
\end{equation}

An interesting feature of the above is that the operators $b_i$ are
linearly dependent on the $a_i$. Thus the components of $\Psi(0^+)$
{\em do not} form a representation of the CCR/CAR's for $M+N$ 
degrees of freedom. In effect the system is reduced to only $M$ 
degrees of freedom, reflecting the fact that the CTC's place 
$N$ constraints on the system. Accordingly, the evolution between
$t=0^-$ and $t=0^+$ is nonunitary, although unitarity is restored
at $t=T^+$. In addition, we see that it is not legitimate to employ
canonical methods to evolve the quantum field in the dischronal region
(if one intends to solve the equation of motion Eq.~(\ref{eq:classf}))
because the data at $t=0^+$ does not obey the CCR/CAR's.

As a final check on our result in this case, and on the loss of degrees
of freedom, let us quantise by the familiar method of obtaining
classical mode solutions. Let
$e_i(t)$ (respectively, $f_j(t)$) be the classical solution to the
free equation of motion with initial data $e_i(0^-)=e_i$
($f_j(0^-)=f_j$) where the basis vectors $e_i$ and $f_j$ were defined in
Sect.~\ref{sect:prelim}. We write the quantum field $\Psi(t)$ as
\begin{equation}
\Psi(t) =  a_i e_i(t) + b_j f_j(t) ,
\end{equation}
where the $a_i$ and $b_j$ form a representation of the CCR/CAR's
on Hilbert space $\frak F$. The components $\Psi_\alpha$ of the field
are obtained by taking the inner product with $v_\alpha$. The time
evolution of the $a_i$ and $b_i$ is defined by re-expressing the field as
\begin{equation}
\Psi(t) = a_i(t) e_i + b_j(t) f_j,
\end{equation}
which leads quickly to the above unitary evolution from $0^-$ to $T^+$ 
using the results of Section~\ref{sect:class}. In the dischronal
region, however, $f_j(t)$ vanishes and so $\Psi(t) = a_i e_i(t)$ and
the reduction to $M$ degrees of freedom is explicit.

\subsection{Interacting Fields}
\label{sect:inter}

Here, we consider three simple examples. Model 1 is a system
with two spatial points and yields a unitary theory for both Fermi
and Bose statistics. Model 2 is a system with three spatial
points. We study this theory for Fermi statistics and show that
the resulting theory is {\em nonunitary}. For simplicity we work in the
appropriate Fock representations and take $A$ and $B$ to be the identity. 

{\noindent\bf Model 1} Our set of spatial points is ${\cal
S}=\{z_1,z_2\}$, and ${\cal S}_i=\{z_i\}$ for $i=1,2$. Thus $W$ is a
$2\times 2$ matrix and $P,Q,R,S$ are scalars. 

{\noindent \em Fermi statistics} The Hilbert space ${\frak F}$
for two fermionic degrees of freedom is isomorphic to ${\Bbb C}^2\otimes
{\Bbb C}^2$. The unique solution to Eq.~(\ref{eq:b}) is 
\begin{equation}
b = (1-S)^{-1}R a ,
\end{equation}
as is easily verified using the fact that $e^{i\kappa a^\dagger a}a=a$. 
Substituting back, we obtain 
\begin{equation}
a(T) = \left(P+Q(1-S)^{-1}R\right)a ,
\end{equation}
which is identical to the unitary free evolution obtained in the previous
subsection. This contrasts with the generically nonunitary evolution
obtained by Politzer~\cite{Pol2} for this model using the
self-consistent path integral -- see Section~\ref{sect:scpi}. 

{\noindent\em Bose Statistics} Here, ${\frak F} = \ell^2
\otimes \ell^2$ (where $\ell^2$ is the Hilbert space of square summable
sequences) and the unique solution to Eq.~(\ref{eq:b}) takes the form 
\begin{equation}
b =  f(a^\dagger a) a ,
\end{equation}
where $f:{\Bbb N}\rightarrow {\Bbb C}$ is defined recursively by
\begin{equation}
f(n+1) = (e^{i\lambda T(n+1) (1+|f(n)|^2)} -S)^{-1}R ,
\end{equation}
with $f(0) = (1-S)^{-1}R$. 

Thus the evolution of $a$ is given by
\begin{eqnarray}
a(T) &=& e^{-i\lambda T a^\dagger 
(1+|f(a^\dagger a)|^2)a}\left(P+Q f(a^\dagger a)\right)a \nonumber \\
&=&  e^{-ig(a^\dagger a)} \left(P+Q(e^{i g(a^\dagger a)}
-S)^{-1}R\right)a ,
\end{eqnarray}
where $g$ is a real-valued function on ${\Bbb N}$ defined by $g(0)=0$
and $g(n) = \lambda T n(1+|f(n-1)|^2)$ for $n\ge 1$. This may be
rewritten as 
\begin{equation}
a(T) = X^\dagger a X ,
\end{equation}
with $X= e^{-i h(a^\dagger a)}$ and $h(n)$  
defined by $h(0)=0$ and
\begin{eqnarray}
e^{-i h(n+1)} &=& e^{-i h(n)} e^{-i g(n)} \nonumber \\
&&\times
\left[P + Q\left(e^{ig(n)}-S\right)^{-1}R\right] .
\end{eqnarray}
The left hand side is always of unit modulus, so $h(n)$
is real-valued and the operator $X$ is unitary. Thus the evolution 
from $t=0^-$ to $t=T^+$ is again unitary. We note that this theory
agrees with the corresponding free theory on ${\frak F}_1$ (though the
theories differ on ${\frak F}_r$ for $r\ge 2$).

{\noindent\bf Model 2}
In this example, our set of spatial points ${\cal S} = \{z_1,z_2,z_3\}$,
is partitioned into ${\cal S}_1 = \{z_1,z_2\}$ and ${\cal S}_2= \{z_3\}$.
The matrix $W$ is now a $3\times 3$ self-adjoint, positive matrix, and
the block decomposition of $e^{-iWT}$ yields a $2\times 2$ matrix $P$,
a 2-dimensional column vector $Q=(Q_1,Q_2)^T$, a 2-dimensional row
vector $R=(R_1,R_2)$ and a scalar $S$.

{\noindent \em Fermi statistics} The Fock space is ${\frak F} = 
{\Bbb C}^2\otimes{\Bbb C}^2\otimes{\Bbb C}^2$. Given operators $a_1$ and
$a_2$ at $t=0$, we seek an operator $b$ such that
\begin{equation}
b = e^{-i\lambda T(a_1^\dagger a_1+a_2^\dagger a_2 +b^\dagger b)}
\left(R_1 a_1 + R_2 a_2 + S b\right) .
\end{equation}

Using the results above, the unique solution to this equation is
\begin{equation}
b = (e^{i\lambda Ta_k^\dagger a_k}-S)^{-1}
(R_1 a_1 + R_2 a_2) ,
\label{eq:c}
\end{equation}
as may easily be checked by decomposing ${\frak F}={\frak
F}_0\oplus{\frak F}_1\oplus{\frak F}_2$ with ${\frak F}_r$ the
eigenspace of $a_k^\dagger a_k$ with eigenvalue $r$. 

\widetext
Substituting, we find that
\begin{equation}
\left(\begin{array}{c} a_1(T) \\ a_2(T) \end{array}\right) =
e^{-i\lambda T(a_k^\dagger a_k+ b^\dagger b)} 
 \left[ P + Q( e^{i\lambda Ta_k^\dagger a_k} - S )^{-1} R
\right]
\left(\begin{array}{c} a_1 \\ a_2 \end{array}\right).
\end{equation}

In Appendix~\ref{sect:detail}, we show that 
\begin{equation}
\langle 0\mid (a_2(T)a_1(T) + a_1(T)a_2(T))a_1^{\dagger}a_2^{\dagger}
\mid 0\rangle 
=F\left(|Q_2|^2-|Q_1|^2\right) \det P , 
\label{eq:J1}
\end{equation}
and
\begin{equation}
\langle 0\mid a_1(T)^2 a_1^{\dagger}a_2^{\dagger}
\mid 0\rangle = F Q_1 \overline{Q_2} \det P ,
\label{eq:K1}
\end{equation}
where
\begin{equation}
F =e^{-i\lambda T}\left\{ \frac{1}{\overline{S}} \left[
\frac{1}{e^{i\lambda T} - S} - \frac{e^{-i\lambda\alpha(S)T}}{1-S}
\right]  + \frac{e^{-i\lambda\alpha(S)T} - 1}{1 - |S|^2} \right\},
\label{eq:F}
\end{equation}
and $\alpha(S)$ is defined by
\begin{equation}
\alpha(S) = \frac{|R_1|^2 + |R_2|^2}{|1 - S|^2} = {1 - |S|^2 \over
|1-S|^2}. 
\label{A}
\end{equation}
Thus, except in the free case or for very carefully tuned parameters the
CAR's are necessarily violated and the evolution is therefore nonunitary.
Note that the coefficients of $F$ in Eqs.~(\ref{eq:J1})
and~(\ref{eq:K1}) vanish simultaneously for all $T$ if and only if $W$
is block diagonal with respect to the decomposition ${\frak H}={\frak
H}_1 \oplus {\frak H}_2$ (in which case $|S|=1$ and
we are no longer in the generic case). 

In the next section, it will be useful to have an explicit matrix
representation for this evolution. Choosing a basis for the
8-dimensional Fock space $\frak F$ such that
\begin{equation}
a_1(0)= \left(\begin{array}{cc} \mbox{\Huge 0} 
& \begin{array}{cc} 1 & 0 \\ 0 & 1 \end{array} \\
\mbox{\Huge 0} & \mbox{\Huge 0} \end{array}\right)\otimes \openone_2,
\qquad\qquad
a_2(0)= \left(\begin{array}{cc} 
\begin{array}{cc} 0 & 1 \\ 0 & 0 \end{array} & \mbox{\Huge 0} \\
\mbox{\Huge 0} & \begin{array}{cc} 0 & -1 
 \\ 0 & 0 \end{array} \end{array}\right)\otimes \openone_2 ,
\end{equation}
where $\openone_2$ is the $2\times 2$ identity matrix, one may show that
\begin{equation}
a_i(T) = \left(\begin{array}{cc} 
\begin{array}{ccc} 0 & M_{i2} &  M_{i1} \end{array} & 0 \\
\mbox{\Huge\bf 0} & 
\begin{array}{c} N_{i1} \\ -N_{i2} \\ 0 \end{array}
\end{array}\right)\otimes \openone_2 .
\end{equation}
Here $M_{ij}$ is the unitary matrix $M=P+Q(1-S)^{-1}R$ and $N$ is also 
a $2\times 2$ unitary matrix given by $N=(P+Q(e^{i\lambda
T}-S)^{-1}R)U$, where $U$ is another $2\times 2$ unitary defined by
\begin{equation}
U_{ij}a_j|_{{\frak F}_2} = 
e^{-i\lambda T(a_k^\dagger a_k+b^\dagger b)}a_i|_{{\frak F}_2} ,
\end{equation}
(which makes sense because the exponential preserves ${\frak F}_1$). The 
precise form of $N$ will not concern us;
however, we note that $M\not=N$, because $a_1(T)$ and $a_2(T)$ fail to
anticommute. 

{\noindent \em Bose statistics} The Fock space is
$\ell^2\otimes\ell^2\otimes\ell^2$ and the unique solution to
Eq.~(\ref{eq:b}) is 
\begin{equation}
b=f(d^\dagger d,c^\dagger c) c ,
\end{equation}
where 
\widetext
\begin{equation}
c=\|R\|^{-1}(R_1 a_1+R_2 a_2), \qquad d=\|R\|^{-1}(R_2 a_1-R_1 a_2),
\label{eq:cd}
\end{equation}
and $f(m,n)$ satisfies
\begin{equation}
f(m,n+1)=\left( e^{i\lambda T[m+(n+1)(1+|f(m,n)|^2)]}-S\right)^{-1}\|R\|,
\end{equation}
with $f(m,0)=(e^{i\lambda T m}-S)^{-1}\|R\|$. Substituting back to determine $a_i(T)$,
we show in Appendix~\ref{sect:detail} that 
\begin{equation}
\langle 0\mid (a_2(T)a_1(T) - a_1(T)a_2(T))d^{\dagger}c^{\dagger}
\mid 0\rangle = F\left(|Q_1|^2+|Q_2|^2\right) \det P  ,
\label{eq:J2}
\end{equation}
with $F$ given by Eq.~(\ref{eq:F}). This should be compared with
Eq.~(\ref{eq:J1}). Thus the evolution fails to be unitary on ${\frak
F}_2$. 

\narrowtext

\section{Discussion of the Nonunitary Evolution}
\label{sect:dnue}

In the previous section, we showed that Model 2 was subject to a
nonunitary evolution for both Bose and Fermi statistics. In this
section, we discuss this evolution in more depth in the fermionic case.
Recall that the Fock space ${\frak F}$ is 8-dimensional, and that the
operators $\Psi_\alpha(0^-)$ ($\alpha=1,2,3$) represent the CAR's for
three degrees of freedom on ${\frak F}$. Writing $\Psi_{1,i}$ for the
operators associated with points in ${\cal S}_1$, and $\Psi_2$ for the
operator associated with the single element of ${\cal S}_2$, we write    
$\Psi_{1,i}(0^-)=a_i$ for $i=1,2$. The Heisenberg evolution
$\Psi_\alpha(0^-)\rightarrow \Psi_\alpha(T^+)$ is such that
$\Psi_{1,i}(T^+)=a_i(T)$ and $\Psi_2(T^+)= \Psi_2(T^-)$.  
Our principal results in this section are, firstly, that the Heisenberg 
picture evolution cannot be expressed in either of the forms 
\begin{equation}
\Psi_\alpha(T^+)=X^{-1}\Psi_\alpha(0^-)X,
\label{eq:frm1}
\end{equation}
or
\begin{equation}
\Psi_\alpha(T^+)=X^\dagger\Psi_\alpha(0^-)X,
\label{eq:frm2}
\end{equation}
for some operator $X$ on $\frak F$; secondly, that the Heisenberg
picture evolution does not admit an equivalent Schr\"{o}dinger picture
description in terms of a superscattering operator. In addition, we will
discuss the problem of extending the evolution from that of the
$\Psi_\alpha(0^-)$ to arbitrary operators on $\frak F$. 

Firstly, then, we show that the Heisenberg picture evolution cannot be
expressed in either of the forms Eq.~(\ref{eq:frm1}) or~(\ref{eq:frm2}).
The form Eq.~(\ref{eq:frm1}) is clearly impossible because it would entail
$\{a_1(T),a_2(T)\}=0$, and we may dispose of
Eq.~(\ref{eq:frm2}) as follows. The explicit form of the
$a_i(T)$ given above shows that any such operator $X$ would necessarily
preserve the subspaces ${\frak F}_0, {\frak F}_1$ and ${\frak F}_2$ of
$\frak F$; moreover, because 
\begin{equation}
a_i(T)|_{{\frak F}_1} = M_{ij}a_j|_{{\frak F}_1},
\end{equation}
where $M$ is unitary, we conclude that $X|_{{\frak F}_1}$ is unitary up to
scale. Then it suffices to note that
\begin{eqnarray}
\{a_1(T),a_2(T)\}\mid 11\rangle & =& X^\dagger (a_1 X X^\dagger 
a_2 \nonumber \\
&& + a_2 XX^\dagger a_1) X \mid 11\rangle
\end{eqnarray}
which vanishes because $X$ preserves ${\frak F}_2$ and $X|_{{\frak
F}_1}$ is unitary up to scale.
Accordingly, we cannot cast the evolution into either of the special
forms Eq.~(\ref{eq:frm1}) or~(\ref{eq:frm2}).

Secondly, we show that the Heisenberg picture evolution cannot be
described by a superscattering operator. Recall that a superscattering
operator on the state space of a (separable) Hilbert space $\frak F$ is 
a linear mapping $\$$ of the trace class operators ${\cal T}({\frak F})$ 
on $\frak F$ such that if $\rho\in{\cal T}({\frak F})$ is a positive
operator\footnote{For our purposes, a ``positive operator'' means
one which is non-negative definite.} of unit trace, then
$\$\rho$ is also a positive element of ${\cal T}({\frak F})$ with unit 
trace. Thus, $\$$ is a linear mapping of density matrices to density
matrices, which need not preserve purity. If a superscattering operator 
$\$$ describes the
Schr\"{o}dinger picture evolution of a system, then the Heisenberg
picture evolution is given by the linear mapping $\$'$ of the bounded
operators ${\frak L}({\frak F})$ on $\frak F$, defined by
\begin{equation}
{\rm Tr}\, \rho (\$' Z)={\rm Tr}\, (\$ \rho) Z ,
\end{equation}
for all $\rho\in{\cal T}({\frak F})$ and $Z\in{\frak L}({\frak F})$. In
fact, $\$'$ is the dual mapping to $\$$ under the natural identification 
of ${\frak L}({\frak F})$ with the dual space of ${\cal T}({\frak F})$.

The dual mapping $\$'$ possesses three easily established properties:
(i) $\$'\openone = \openone$; (ii) $(\$' Z)^\dagger = \$'(Z^\dagger)$
for all $Z$; and (iii) $\$'$ is positive in the sense that $\$' Z$ is a  
positive operator whenever $Z$ is. If one writes the superscattering
operator using the index notation (e.g.,~\cite{Hawk}) 
\begin{equation}
(\$\rho){}^A{}_B = \${}^A{}_{BC}{}^D \rho{}^C{}_D ,
\end{equation}
then $\$'$ may be written as
\begin{equation}
(\$'Z){}^D{}_C = \${}^A{}_{BC}{}^D Z{}^B{}_A .
\label{eq:dollarc}
\end{equation}

Returning to our case of interest, we now show that there is no
superscattering operator $\$$ for which the Heisenberg
evolution can be written as
\begin{equation}
\Psi_\alpha(T^+)=\$'\Psi_\alpha(0^-) .
\label{eq:sufrm}
\end{equation}
We will need the fact that if $u=(u_1, u_2)^T$ and
$v=(v_1, v_2)^T$ are 2-dimensional complex column vectors and 
$\alpha\in{\Bbb R}$, then the eigenvalues $\mu$ of the matrix
\begin{equation}
K = \left(\begin{array}{cccc} \alpha & u_2 & u_1 & 0 \\
\overline{u_2} & \alpha & 0 & v_1 \\ \overline{u_1} & 0 & \alpha & -v_2 \\
0 & \overline{v_1} & -\overline{v_2} & \alpha
\end{array}\right) \otimes \openone_2 ,
\label{eq:mafrm}
\end{equation}
satisfy
\begin{equation}
(\mu-\alpha)^4 - (\mu-\alpha)^2 (u^\dagger u+ v^\dagger v) + 
| u^\dagger v|^2 = 0.
\end{equation}
Note that the operator $C=\alpha\openone+\beta
a_1+\overline{\beta} a_1^\dagger + \gamma
a_2+\overline{\gamma}a_2^\dagger$ takes the form~(\ref{eq:mafrm}) with
$u=v=(\beta, \gamma)^T$ and therefore has eigenvalues 
\begin{equation}
\mu=\alpha\pm\sqrt{|\beta|^2+|\gamma|^2} .
\end{equation}
Accordingly, $C$ is positive if and only if
$\alpha\ge\sqrt{|\beta|^2+|\gamma|^2}$. Now suppose that there exists a 
superscattering operator $\$$ such that Eq.~(\ref{eq:sufrm}) holds.
Then using the properties (i) and (ii) of $\$'$, we have
\begin{eqnarray}
\$' C & = & \alpha \openone + \beta
a_1(T)+\overline{\beta} a_1(T)^\dagger \nonumber \\
&& + \gamma a_2(T)+\overline{\gamma}a_2(T)^\dagger ,
\end{eqnarray}
which may be seen to take the form~(\ref{eq:mafrm}) with 
$u=((\beta, \gamma)M)^T$, $v=((\beta, \gamma)N)^T$. Hence
its eigenvalues are
\begin{equation}
\mu = \alpha\pm\sqrt{|\beta|^2+|\gamma|^2\pm \Delta} ,
\end{equation}
where the two $\pm$ signs are independent and 
\begin{equation}
\Delta = \sqrt{(|\beta|^2+|\gamma|^2)^2-|u^\dagger v|^2}
\end{equation}
is real and positive by the Schwarz inequality and the
unitarity of $M$ and $N$. Because these matrices are unequal,
we may choose $\beta$ and $\gamma$ so that $\Delta>0$. 
Choosing $\alpha$ to lie in the range
\begin{equation}
\sqrt{|\beta|^2+|\gamma|^2}< \alpha <
\sqrt{|\beta|^2+|\gamma|^2+\Delta}, 
\end{equation}
the operator $C$ is positive, but $\$' C$ is not. Accordingly,
$\$'$ violates property (iii) above and therefore
cannot be the dual of a superscattering operator.

Next, we consider the Heisenberg evolution itself in more detail. 
It is worth pointing out that we have not by any means obtained the full 
Heisenberg picture evolution; at present we know the evolution of only a  
3-dimensional subspace (spanned by the $\Psi_\alpha(0^-)$) of the
64-dimensional space ${\frak L}({\frak F})$ of linear operators on the
8-dimensional Hilbert space ${\frak F}$. Owing to our results above,
various natural strategies for extending this evolution to the whole of
${\frak L}({\frak F})$ are denied to us: the evolution cannot be
extended as a $*$-homomorphism (i.e., mapping any polynomial in the
$\Psi_\alpha(0^-)$ to the corresponding polynomial in the
$\Psi_\alpha(T^+)$) because the CAR's are violated; we cannot
write $Z\rightarrow X^{-1}Z X$ or $Z\rightarrow X^\dagger ZX$ because of
our observations above, nor can we write $Z\rightarrow \$'Z$ for some
superscattering operator $\$$. 

It therefore seems that there is no natural extension of our evolution
to ${\frak L}({\frak F})$. As a concrete illustration of this type of
behaviour, let us consider an example with one fermionic degree of
freedom. Define the operator $a$ on ${\Bbb C}^2$ by
\begin{equation}
a = \left(\begin{array}{cc} 0 & 1 \\ 0 & 0 \end{array} \right) , 
\end{equation}
and suppose an evolution is given such that $\openone\rightarrow \openone$,
$a\rightarrow \mu a$ and $a^\dagger\rightarrow\mu a^\dagger$, where
$0\le \mu < 1$. It turns out that there are at least two choices for the
evolution of $a^\dagger a$ consistent with a superscattering operator
description. The first is that 
$a^\dagger a\rightarrow a^\dagger a$, corresponding to a superscattering
operator $\$$ with action
\begin{equation}
\$ \left(\begin{array}{cc} \alpha & \beta \\ \overline{\beta} & 1-\alpha
\end{array}\right)
=\left(\begin{array}{cc} \alpha & \mu\beta \\ \mu\overline{\beta} & 1-\alpha
\end{array}\right) ,
\end{equation}
on the state space of ${\Bbb C}^2$,
whilst the second is $a^\dagger a \rightarrow aa^\dagger$ and corresponds
to the superscattering operator $\pounds$ with action
\begin{equation}
\pounds \left(\begin{array}{cc} \alpha & \beta \\ \overline{\beta} & 1-\alpha
\end{array}\right)=
\left(\begin{array}{cc} 1-\alpha & \mu\beta \\ \mu\overline{\beta} & \alpha
\end{array}\right).
\end{equation} 

To conclude this section, we note that the failure of positivity which
showed the nonexistence of a superscattering operator can be traded for
a loss of the trace preserving property: by allowing $\openone
\rightarrow \kappa \openone$ for $\kappa \ge \sqrt{2}$, any positive $C$ 
of the form discussed above is mapped to a positive operator. One
might therefore attempt to extend this in some way to a positive
evolution on the
whole of ${\frak L}({\frak F})$ (which can be done if the evolution on
$\openone, a_i(0), a_i(0)^\dagger$ is {\em completely positive} -- see
Theorem~1.2.3 in Arveson~\cite{Arves}) thereby obtaining (by duality) a
Schr\"{o}dinger picture evolution possessing all the properties of a
superscattering matrix except the preservation of trace. Rather than
allowing individual probabilities to be negative with total probability
equal to unity, we would now have positive probabilities with a 
total in excess of unity. It would be tempting to rescale this total to
remove this problem, but that would amount to rescaling $a_i(T)$,
for which there is no obvious justification. 

\section{The Classical Limit}
\label{sect:classlim}

With the normal ordering used above, we have shown that the quantum
theory is uniquely determined in the generic case for all values of the
coupling constant $\lambda$. On the other hand, we have also seen that
the classical theory is nonunique in the strong coupling regime. It is
therefore interesting to determine the extent to which the classical
theory may be regarded as a limit of the quantum theory.

We consider Model 1 with Bose statistics. Reintroducing the units of
action by replacing $a$ and $b$ by $\hbar^{1/2} a$ and $\hbar^{1/2} b$
respectively, the consistency requirement Eq.~(\ref{eq:b}) becomes
\begin{equation}
b = e^{-i\hbar\lambda T(a^\dagger a + b^\dagger b)}\left(Ra+Sb\right) ,
\end{equation}
in which $a$ and $a^\dagger$ obey the CCR's $[a,a^\dagger]=1$. The
unique solution to this is $b= f(\hbar\lambda T; a^\dagger a)a $, where 
\begin{equation}
f(\nu;n+1) = (e^{i\nu (n+1)(1+|f(\nu;n)|^2)}-S)^{-1} R,
\end{equation}
with $f(\nu;0)=(1-S)^{-1}R$ for all $\nu$. 

One expects the classical limit to be
\begin{equation}
b_{\rm cl} = g(\lambda T |a_{\rm cl}|^2)a_{\rm cl} ,
\end{equation}
where
\begin{equation}
g(\mu)= \lim_{n\rightarrow\infty} {f}(\mu/n;n) ,
\end{equation}
which implements the limit $\hbar\rightarrow 0$ while keeping $\lambda T\hbar
n=\mu$ fixed, where we define $\mu=\lambda T |a_{\rm cl}|^2$. Note that the
classical quantities $a_{\rm cl}$
and $b_{\rm cl}$ have dimensions of $({\rm action})^{1/2}$. If the limit
$g(\mu)$ exists uniformly in $\mu$ in some neighbourhood,
then we have
\begin{equation}
b_{\rm cl} = \left(e^{i\lambda T (|a_{\rm cl}|^2+|b_{\rm
cl}|^2)}-S\right)^{-1}R ,
\end{equation}
for initial data corresponding to this neighbourhood. This is, of course,
equivalent to the classical consistency condition
Eq.~(\ref{eq:bcon}) for this model. 

Thus it is important to clarify the convergence properties of 
$g_n(\mu)={f}(\mu/n;n)$ as $n\rightarrow\infty$. This turns out to
be somewhat delicate and we have not as yet brought the general case
fully under analytic control. However, we have investigated the problem
numerically in the particular example $P=-Q=R=S=1/\sqrt{2}$ and $A=B=1$
for various values of $\mu$. With these parameters, it is easy to see
that $(g_n(\mu)^{-1}+1)/\sqrt{2}$ lies on the unit circle for all
$n,\mu$; accordingly we define 
\begin{equation}
\theta_n=\arg \frac{g_n(\mu)^{-1}+1}{\sqrt{2}} ,
\label{eq:thndf}
\end{equation}
where
$-\pi<\arg z \le \pi$. In Table~\ref{table1}, we give the values of
$\theta_n$ for $n=500$, $1000$, $1500$, $2000$ and $2500$ with
$\mu=0.5$, $3.0$, $7.0$, and $13.0$ and also give the value of $\theta$ 
corresponding to the nearest solution to the classical
equation of motion. These results suggest a slow convergence to a unique 
classical limit for each of these values of $\mu$. At these values of
$\mu$, the classical system has $1$, $7$, $13$ and $25$  
solutions respectively (see Figures~\ref{fig1} and~\ref{fig2} for the
cases $\mu=0.5$ and $\mu=3.0$) of
which the quantum theory picks out precisely one. However, it is by no
means the case that $\theta_n$ converges for any value of $\mu$. Indeed,
it appears that convergence occurs only for certain bands of $\mu$,
the first two of which are $0\le \mu\lesssim 3.5$ and $6.7\lesssim \mu
\lesssim 8.5$. We believe that this band structure continues
indefinitely; it seems reasonable to assume that the bands become narrower
as $\mu$ increases.  

One may interpret these results as indicating that for those ranges of
the coupling strength $\mu=\lambda T|a_{\rm cl}|^2$ in which the
classical limit exists, the unique classical solution isolated by the
quantum theory is the correct classical description of the dynamics.
However, for other values of the coupling strength there is no classical
limit and presumably no valid classical description of the dynamics.
Finally, we note that these results are strongly dependent on the choice
of operator ordering as we will see in the next section.

\section{Operator Ordering}
\label{sect:fact}

So far, we have worked with a single choice of operator ordering, namely
the literal ordering of Eq.~(\ref{eq:classf}) which corresponds to
normal ordering of the quantised Hamiltonian. In this section, we
briefly discuss the effect of allowing alternative orderings in which
Eq.~(\ref{eq:classf}) is ordered as
\begin{equation}
\dot{\psi} = -iW\psi - i\alpha\lambda(\psi^\dagger\psi)\psi -
i(1-\alpha)\lambda \psi(\psi^\dagger\psi) ,
\label{eq:reorder}
\end{equation}
for $\alpha\in [0,1]$. The foregoing treatment is the case $\alpha=1$. 

Consider Model 1 for Bose statistics. The analogue of
Eq.~(\ref{eq:b}) is 
\begin{equation}
b = e^{-i\alpha\lambda T(a^\dagger a+b^\dagger b)}
(Ra + Sb) e^{-i(1-\alpha)\lambda T(a^\dagger a+b^\dagger b)},
\end{equation}
(we have set $A=\openone$ for simplicity). Making the {\em ansatz} $b=
f(a^\dagger a)a$, we find that $f$ satisfies
\begin{equation}
f(0)= \left( R+Sf(0)\right) e^{-i\lambda T (1-\alpha)(1+|f(0)|^2)},
\end{equation}
and
\begin{eqnarray}
f(n+1) &=& e^{-i\lambda T (n+1) \alpha(1+|f(n)|^2)}\left( R+Sf(n+1)
\right) \nonumber \\
&& \times
e^{-i\lambda T (n+2) (1- \alpha)(1+|f(n+1)|^2)},
\end{eqnarray}
for $n\ge 0$. The case $\alpha=1$ was treated in Section~\ref{sect:canon}
and uniquely determines $f(n+1)$ in terms of $f(n)$ for each $n$.
However, the case $\alpha=0$ is rather different and is described by
\begin{equation}
f(n) = e^{-i\lambda T\hbar(n+1)(1+|f(n)|^2)}\left( R+Sf(n)\right),
\label{eq:fa0}
\end{equation}
where we have written $\hbar$ explicitly. It is easy to recast this into
the form of the {\em classical} consistency requirement
Eq.~(\ref{eq:bcon}) and it follows that $f(n)$ is uniquely
determined for small quantum numbers $n\hbar \ll (\lambda T)^{-1}$ but
not for $n\hbar\gg(\lambda T)^{-1}$, i.e., classical nonuniqueness
reemerges at high quantum numbers. There are therefore many functions
$f(n)$ solving Eq.~(\ref{eq:fa0}), each one of which corresponds to a
different ``branch'' of the quantum theory. Most of these branches do
not possess a classical limit. However, in contrast to
the situation for normal ordering, {\em every} classical solution will
arise as the classical limit of some branch of the quantum theory. 

It would be interesting if the nonunitarity of Model 2 could be removed 
by a suitable ordering prescription. In Appendix~\ref{sect:detail}, we
investigate this for  orderings of form~(\ref{eq:reorder}) with
the {\em ansatz} $b=f(d^\dagger d,c^\dagger c)c$ with $c$ and $d$ given
by Eq.~(\ref{eq:cd}). (Solutions do exist
which take this form.) For all $0\le \alpha\le 1$, we find that the
(anti)commutation relations are violated for generic values of the
parameters.

\section{Self-Consistent Path Integral}
\label{sect:scpi}
\subsection{General Formalism}
\label{sect:scpigf}

In this section, we compare the results obtained from the QIVP with
those obtained using the self-consistent path integral developed by
Thorne and collaborators~\cite{FMNEKTY,EKT,KT} and employed by 
Politzer~\cite{Pol2}. To
establish our notation, we briefly review the quantisation of our system 
by path integral methods in the absence of CTC's. Starting with the
bosonic case, it is convenient to use the holomorphic representation
(see, e.g., \cite{FS}) in which the Hilbert space
${\frak F}$ is the space of analytic functions
$f(\overline{c_1},\ldots,\overline{c_s})$  with inner product
\begin{equation}
\langle f\mid g\rangle = 
\int {\cal D}c^\dagger{\cal D}c \,e^{-c^\dagger c}
\overline{f(c^\dagger)} 
g(c^\dagger) ,
\end{equation}
where we write $c^\dagger$ to denote
$(\overline{c_1},\ldots,\overline{c_s})$ and the
measure is
\begin{equation}
{\cal D}c^\dagger{\cal D}c =\prod_j \frac{d\overline{c_j}dc_j}{2\pi i} .
\end{equation}

The Hilbert space ${\frak F}$ carries a (Fock) representation of the
CCR's in which $c_j^\dagger$ acts as multiplication by $\overline{c_j}$ 
and $c_j$ as $\partial/\partial\overline{c_j}$. Operators on $\frak F$
are described by  their kernels:
\begin{equation}
(Af)(c^\dagger) = \int {\cal D}{c^\prime}^\dagger{\cal D}c^\prime \,
e^{-{c^\prime}^\dagger c^\prime}
A(c^\dagger;c^\prime)f({c^\prime}^\dagger) .
\end{equation}
In particular, if $K$ is a $s\times s$ matrix, then the mapping
$f(c^\dagger)\rightarrow f(c^\dagger K)$ has kernel $\exp c^\dagger
Kc'$. 

\widetext
Starting with the (normal ordered) quantised bosonic Hamiltonian $H$ on 
${\frak F}$, one may obtain the kernel for $U=e^{-iHt}$ in the form
\begin{equation}
U_t(c^\dagger;c^\prime) = 
\int \prod_{t'}
{\cal D}\gamma(t')^\dagger{\cal D}\gamma(t') \, 
\exp\left\{\frac{1}{2}(\gamma^\dagger(t)\gamma(t)+
\gamma^\dagger(0)\gamma(0)) + i S[\gamma]\right\} ,
\label{eq:PIprop}
\end{equation}
where the action functional $S[\gamma]$ is defined in terms of the
classical Hamiltonian~(\ref{eq:Ham1}) by
\begin{equation}
S[\gamma] = \int_0^t 
\left(\frac{i}{2}(\gamma(t')^\dagger\dot{\gamma}(t')
- \dot{\gamma}(t')^\dagger \gamma(t'))
- H(\gamma(t'),i\gamma(t')^\dagger) \right) dt' ,
\end{equation}
\narrowtext
and the paths $\gamma(t')$ are
subject to the boundary conditions $\gamma^\dagger(t) =c^\dagger$ and
$\gamma(0)=c^\prime$. In the free case, for example, one may evaluate
the path integral explicitly to give
\begin{equation}
U_t(c^\dagger;c^\prime)= \exp c^\dagger e^{-iWt} c^\prime.
\end{equation}

One may develop the path integral treatment for Fermi statistics
in a parallel fashion~\cite{FS} by replacing the integration variables
by Grassmann numbers and ${\cal D}c{\cal D}c^\dagger$ by Berezin measure.
Again, the resulting kernel has the action of $e^{-iHt}$ on ${\frak F}$,
where $H$ is now the fermionic normal ordered quantised Hamiltonian.

A natural generalisation of this to enable the treatment of  
chronology violating systems is the {\em self-consistent path
integral}~\cite{Pol2,FMNEKTY,EKT,KT}. Instead of integrating over all
field configurations with $\gamma(0)=c'$ and
$\gamma^\dagger(T)=c^\dagger$ to form the kernel $U_T(c^\dagger,c')$,
the self-consistent path integral prescription requires that one should
restrict the class of field configurations
to those obeying the self-consistency requirements imposed by any CTC's
present (here, the boundary conditions~(\ref{eq:CTCbcs})). 
To implement this, we first decompose ${\frak F}={\frak F}_1\otimes
{\frak F}_2$, where ${\frak F}_1$ is the space of analytic functions 
in variables $\overline{a_1},\ldots,\overline{a_{s_1}}$, and ${\frak
F}_2$  is the space of analytic functions in $\overline{b_1},\ldots,
\overline{b_{s_2}}$. The (self-consistent) evolution kernel from $t=0^-$  
to $t=T^+$ can then be written in the form 
\begin{equation}
X(a^\dagger,b^\dagger;a',b') = {\cal N} e^{b^\dagger B b'} 
\widetilde{U_T}(a^\dagger;a').
\end{equation}
Here, ${\cal N}$ is a normalisation constant and the factor 
$e^{b^\dagger B b'}$ implements the boundary condition 
$\psi_2(T^+)=B\psi_2(0^-)$ while $\widetilde{U_T}$
is given by the same path integral as $U_T$ but taken over all field 
configurations with $\gamma(0) = (a',b'')$, $\gamma^\dagger(T) = 
(a,A b'')^\dagger$ for any $b''$. As noted by Politzer~\cite{Pol2}, 
$\widetilde{U_T}(a^\dagger;a')$ may be obtained from 
$U_T(a^\dagger,b^\dagger;a',b')$ by setting $b = A b'$ and
integrating over all possibilities in the Hilbert space measure of 
${\frak F}_2$, that is,
\begin{equation}
\widetilde{U_T}(a^\dagger;a^\prime) = 
\int {\cal D}b^\dagger {\cal D}b \, e^{-b^\dagger b} 
U_T(a^\dagger,b^\dagger A^\dagger; a^\prime,b),
\end{equation}
which may be rewritten in the form
\begin{eqnarray}
\widetilde{U_T}(a^\dagger;a^\prime) &=& 
\int {\cal D}b^\dagger {\cal D}b \int {\cal D}c^\dagger {\cal D}c\, 
e^{-b^\dagger b-c^\dagger c} \nonumber \\
&&\times e^{b^\dagger A^\dagger c} 
U_T(a^\dagger,c^\dagger; a^\prime,b) .
\label{eq:4.3}
\end{eqnarray}
Thus, by expanding $e^{b^\dagger A^\dagger c}$ as 
\begin{equation}
e^{b^\dagger A^\dagger c} = \prod_{i=1}^{s_2} \sum_{n_i=0}^\infty
\frac{(A^\dagger c)_i^{n_i} (b_i^\dagger)^{n_i}}{n_i!},
\end{equation}
we obtain the matrix element $\langle{\bf m}\mid \widetilde{U_T}\mid
{\bf m'}\rangle$ in the form
\begin{equation}
\langle {\bf m}\mid \widetilde{U_T}\mid {\bf m^\prime}\rangle
= \sum_{\bf n} \langle {\bf m};{\bf \tilde{n}}
\mid U_T \mid {\bf m^\prime};{\bf n}\rangle ,
\label{eq:ptdf}
\end{equation}
where the vector $\mid{\bf m}\rangle\in{\frak F}_1$ is the function
$\prod_i (m_i!)^{-1/2} \overline{a_i}^{m_i}$, and the vector
$\mid{\bf \tilde{n}}\rangle\in{\frak F}_2$ is the function
$\prod_i (n_i!)^{-1/2} \overline{(A^\dagger b)_i}^{n_i}$.
In addition, $(A^\dagger b)_i$ denotes $\sum_j A^\dagger_{ij}b_j$. 
We refer to Eq.~(\ref{eq:ptdf}), which is a generalisation of the
expression given by Politzer~\cite{Pol2} as the {\em partial trace 
definition} of the self-consistent path integral. 

The fermionic case follows a similar pattern, when one replaces
the integration variables by Grassmann numbers and uses Berezin measure; 
the main difference lies in the partial trace definition. Starting
from the analogue of~(\ref{eq:4.3}), we expand $e^{b^\dagger 
A^\dagger c}$ as
\begin{eqnarray}
e^{b^\dagger A^\dagger c} &=& \prod_{i=1}^{s_2} \left( 1+ b_i^\dagger 
(A^\dagger c)_i\right) \nonumber \\
&=& \sum_{\bf n} (-1)^n\prod_i (A^\dagger
c)_i^{n_i} 
(b_i^\dagger)^{n_i} ,
\end{eqnarray}
where $n=\sum_i n_i$, and therefore obtain
\begin{equation}
\langle {\bf m}\mid \widetilde{U_T}\mid {\bf m^\prime}\rangle
= \sum_{\bf n} (-1)^{n} \langle {\bf m};{\bf \tilde{n}}
\mid U_T \mid {\bf m^\prime};{\bf n}\rangle ,
\label{eq:PTdf2}
\end{equation}
under the assumption that the Grassmann number $b_i^\dagger$ 
commutes with the kernel of $U_T(a^\dagger,c^\dagger;a',b)$, which holds
if $H$ conserves particle number (as it does in our case of interest).
The factor of $(-1)^n$
was omitted by Politzer~\cite{Pol2}; it arises because terms of the
form $b_i^\dagger (A^\dagger c)_i$ coming from $e^{b^\dagger A^\dagger c}$
must be rearranged in order to move the $b_i^\dagger$'s into the ket
and the $(A^\dagger c)_i$'s into the bra of the matrix element
$\langle {\bf m};{\bf \tilde{n}}\mid U_T \mid 
{\bf m^\prime};{\bf n}\rangle$. In Appendix~\ref{sect:PTF}, we will see
how, for free fields, these factors ensure that the evolution computed 
from~(\ref{eq:PTdf2}) agrees with that obtained directly from the path 
integral, and also with that obtained from the QIVP.

\subsection{Free Fields}

Whilst one can use the partial trace definition to compute the quantum
evolution $X$ for free fields (see Appendix~\ref{sect:PTF}), it
is easier to evaluate the path integral directly, using the
fact that the kernel of the free evolution is given by
\begin{equation}
U_T(c^\dagger;c^\prime)= \exp c^\dagger e^{-iWT} c^\prime.
\end{equation}
Writing $e^{-iWT}$ in the block form~(\ref{eq:block}) as above, we
obtain 
\begin{eqnarray}
\widetilde{U_T}(a^\dagger;a^\prime) &=& 
\int {\cal D}b^\dagger {\cal D} b \,\exp\left\{
-b^\dagger(\openone -A^\dagger S) b +a^\dagger P a^\prime \right.
\nonumber \\
&& \left. + a^\dagger Q b + 
b^\dagger A^\dagger R a^\prime \right\} ,
\end{eqnarray}
which may be evaluated to give
\begin{equation}
\widetilde{U_T}(a^\dagger;a^\prime) = 
(\det (\openone - A^\dagger S))^{-1} \exp a^\dagger M a^\prime ,
\end{equation}
where $M=(P+Q(A-S)^{-1}R)$. In the generic case, the convergence of the  
path integral is guaranteed because $\|A^\dagger S\|<1$ and so 
$\openone - A^\dagger S$ has positive hermitian part.

Noting that $V(a^\dagger;a^\prime)=\exp a^\dagger M a^\prime$ is the
unitary kernel, because $M$ is unitary, we conclude that the unitary 
kernel obtained from the self-consistent path integral is
\begin{equation}
X(a^\dagger,b^\dagger;a^\prime,b^\prime) = \exp 
\left\{a^\dagger M a^\prime
+ b^\dagger B b^\prime\right\} ,
\end{equation}
whose corresponding operator $X$ acts on annihilation operators $a_i$
and $b_i$ according to
\begin{equation}
X^\dagger a_i X = M_{ij}a_j,\qquad X^\dagger b_i X = B_{ij}b_j .
\label{eq:Xact}
\end{equation}
Moreover, the normalisation constant is given by ${\cal N}=\det(\openone 
- A^\dagger S)$.

In the fermionic case, the path integral may be evaluated explicitly
to obtain a unitary evolution with the action~(\ref{eq:Xact}) on
annihilation operators and normalisation constant ${\cal N} =
\det(\openone- A^\dagger S)^{-1}$. 

Thus in both cases, we have obtained agreement with the QIVP evolution. 
Moreover, we have given a general proof of the unitarity of free field
evolution using the self-consistent path integral; previously this had
only been established in a particular case~\cite{Pol2}.

\subsection{An Interacting Model}
\label{sect:ipathi}

We study Model 1 of Section~\ref{sect:canon} for both Bose and
Fermi statistics, employing the partial trace definition, and
choosing the normalisation constant so that $\langle  0; 0\mid X\mid  
0;0\rangle =1$, which is reasonable because the Hamiltonian $H$ is
particle-number preserving. In the fermionic case, we obtain
\begin{eqnarray}
\langle 0\mid \widetilde{U_T}\mid 0\rangle &=&
 \langle 00\mid e^{-iHT} \mid 00\rangle
-\langle 01\mid e^{-iHT} \mid 01\rangle \nonumber \\
&=& 1-S
\nonumber\\
\langle 1\mid \widetilde{U_T}\mid 1\rangle &=& 
\langle 10\mid e^{-iHT} \mid 10\rangle
-\langle 11\mid e^{-iHT} \mid 11\rangle \nonumber \\
& =& 
P - (PS-RQ) e^{-i\lambda T} ,
\end{eqnarray}
from which it follows that the evolution from $t=0^-$ to $t=T^+$ is
given by
\begin{equation}
\langle m;n\mid X\mid m;n\rangle = \delta_{nn'}\delta_{mm'} f(m) ,
\end{equation}
where
\begin{equation}
f(m) = \left\{\begin{array}{cl} 1 & m=0 \\
\frac{P-(PS-RQ)e^{-i\lambda T}}{1-S} & m=1. \end{array}\right. 
\end{equation} 
Thus $X$ is nonunitary in general, which is essentially the result
obtained by Politzer~\cite{Pol2} in special cases, modulo some changes
of sign
owing to the factors of $(-1)^n$ discussed above. Except when $\lambda
T/(2\pi)\in{\Bbb Z}$ this differs from the unitary evolution obtained
from the QIVP. 

\widetext
In the bosonic case, we have
\begin{equation}
\langle m\mid \widetilde{U_T} \mid m'\rangle 
= \sum_{n=0}^\infty\frac{e^{-i\lambda T(m+n)(m+n-1)/2}}{(m!n!)^{1/2}}
\langle 00\mid (Ra+Sb)^n(Pa+Qb)^m\mid m'n\rangle ,
\end{equation}
and therefore conclude that $\langle mn\mid X\mid m'n'\rangle
= \delta_{mm'}\delta_{nn'} f(m)$ with $f(0)=1$ and
\begin{equation}
f(m) = \frac{\sum_{n=0}^\infty
e^{-i\lambda T(m+n)(m+n-1)/2}\sum_{r={\rm max}\,\{n-m,0\}}^n
\left(\begin{array}{c} n \\ r \end{array}\right)
\left(\begin{array}{c} m \\ n-r \end{array}\right)
(RQ)^{n-r}P^{m+r-n}S^r}{\sum_n S^n e^{-i\lambda Tn(n-1)/2}} .
\end{equation}
One may show that $X$ fails to be unitary in general. Again, it clearly 
differs from the unitary evolution obtained from the QIVP. 
\narrowtext

\section{Conclusion}
\label{sect:concl}

In this paper, we have analysed in detail the classical and quantum
behaviour of a class of nonlinear chronology violating systems. 
Classically, we found that unique solutions exist for all choices of
initial data in the linear and weak-coupling regimes, whilst the
solutions become nonunique in the strong-coupling regime. This confirms
the expectation that the behaviour of nonlinear fields
interpolates between that of classical linear fields and hard-sphere
mechanics. Quantum mechanically, we have shown that one can make sense
of the quantum initial value problem for chronology violating systems;
moreover, (at least with a natural choice of operator ordering) the
quantum  dynamics is unique for all values of the coupling constant. We
have also exhibited examples in which this evolution does not preserve
the (anti)commutation relations; it seems highly likely that this is the 
general situation. Moreover, the nonunitary evolution cannot be
described by a superscattering operator -- the loss of unitarity is more 
radical than previously thought, e.g., by Hawking~\cite{Hawk}.

We have also compared our quantum evolution with that computed using the
self-consistent path integral, and found that they do not agree. This is
not surprising, because the equivalence of these approaches for
non-chronology violating systems relies on the existence of
a foliation by Cauchy surfaces and there is no {\em a priori} reason to
expect the equivalence to persist in the presence of CTC's.  
In this regard it is interesting
that the QIVP and self-consistent path integral {\em are} nonetheless
equivalent for linear fields. To some extent, it is a matter of
taste which approach one prefers. We prefer the QIVP
approach for two main reasons. Firstly, we have found circumstances
(e.g., Model~1 in Section~\ref{sect:inter}) in which one obtains a
unitary theory from the QIVP but not from the path integral. Secondly,
the effect of the CTC's in our models is to introduce 
constraints which lead to a nontrivial geometric structure in the
classical phase space. We suspect that the quantisation of this system
requires more than just a restriction of the class of allowed histories,  
and that the path integral measure should also be modified
(a similar comment has also been made in~\cite{Arley}). A hint of
this appears in the treatment of linear fields, in which the propagator
obtained from the self-consistent path integral must be rescaled by a
factor of $\det(\openone-A^\dagger S)^{\pm 1}$. It is plausible that in
the linear case, the required modification to the path integral measure
reduces to rescaling by this constant factor, but that for the nonlinear
case the modification is nontrivial. At present it is not clear to us
exactly how the path integral
should be modified; on the other hand it is clear that the QIVP does
correctly implement the CTC constraints and remains close to the spirit
of the classical treatment. 

The relationship between the unique quantum theory and the nonunique
classical theory is intriguing. We have seen that there exist ranges of
the coupling strength in which the quantum theory has a classical limit
which selects precisely one of the many classical solutions, and other
ranges in which no classical limit exists. It would be interesting to
understand the underlying reasons for this behaviour in more detail. 

Finally, it is curious that the classical symplectic structure can be
preserved for systems which do not preserve the quantum commutation
relations. It is tempting to wonder whether there is a way of
quantising these models so that unitarity is preserved. Our
uniqueness result for the QIVP rules this out within a Hilbert space
context (at least with normal operator ordering) but it is
possible that the situation might be different for the QIVP on an
indefinite (Krein) inner product space in which irreducible non-Fock
representations of the CCR's exist for even a single degree of
freedom~\cite{Mnat}. The motivation for studying Krein spaces would
be that the loss of physical degrees of freedom in the nonchronal region 
might be equivalent to the addition of unphysical states with negative
norm-squared.

\acknowledgments

We thank John Friedman for correspondence on the preservation of the
symplectic structure and Simon Eveson, Petr H\'aj\'{\i}{\v c}ek and 
Bernard Kay for
useful and clarifying conversations.  The work of CJF was supported by
the Royal Society, the Schweizerischer Nationalfonds and EPSRC Grant No.
GR/K~29937 to the University of York. The work of AH
was supported in part by the Schweizerischer Nationalfonds and the
Tomalla Foundation at the Universit\"at Bern.

\appendix

\section{Path Integral Approach to the Free Classical Evolution}
\label{sect:FQPI}

In this Appendix, we show how the classical evolution derived in
Section~\ref{sect:freec} may be reproduced using a method due to
Goldwirth {\em et al.} \cite{GPPT} and based on path integrals.
(Goldwirth {\em et al.} regarded the classical wave equation as the
first quantisation of an underlying particle mechanics.) 
The central idea is to sum the propagators
for all possible trajectories through the CTC region. We will
use this method to determine the propagator between $t=0^-$ and
$t=T^+$, essentially repeating the calculation of~\cite{GPPT} in our 
(slightly simpler) notation.

The block matrix decomposition Eq.~(\ref{eq:block}) suggests that
we break the problem into four parts, evaluating the propagators
from ${\cal S}_i$ at $t=0^-$ to ${\cal S}_j$ at $t=T^+$ separately
for each $i,j=1,2$. Note that a particle on ${\cal S}_2$ at $t=0^-$
must enter the wormhole there and reemerge on ${\cal S}_2$ at $t=T^+$.
Thus the ${\cal S}_2\rightarrow {\cal S}_2$ propagator equals $B$,
whilst 
that for ${\cal S}_2\rightarrow {\cal S}_1$ vanishes. In addition,
the propagator ${\cal S}_1\rightarrow {\cal S}_2$ also vanishes by the 
time reverse of this argument. It remains to compute the propagator
for ${\cal S}_1\rightarrow {\cal S}_1$. In this case, there are
countably  many
possible trajectories. The particle can either go directly to ${\cal
S}_1$ 
with propagator $P$, or it can enter the CTC region to
arrive at ${\cal S}_2$ at $t=T^-$ (propagator $R$), pass through the
wormhole to ${\cal S}_2$ at $t=0^+$ (propagator $A^{-1}$), execute
$n$ circuits of the CTC's (propagator $(A^{-1}S)^n$) and finally
travel from ${\cal S}_2$ at $t=0^+$ to ${\cal S}_1$ at $t=T^+$
(propagator $Q$). The combined propagator for this trajectory is
$Q(A^{-1}S)^n A^{-1}R$; summing over all possible winding numbers and
the direct trajectory, we obtain the total propagator
\begin{eqnarray}
M &=& P + Q\left(\sum_{n=0}^\infty (A^{-1}S)^n \right)A^{-1}R 
\nonumber \\  
&=& P + Q(A-S)^{-1}R ,
\end{eqnarray}
which agrees with the result obtained in Section~\ref{sect:freec}.

\widetext

\section{Partial Trace Formalism for Free Fields}
\label{sect:PTF}

In this Appendix, we derive the evolution operator for free field
models in the presence of CTC's using the partial trace formulation
of the self-consistent path integral.

We consider a general free theory whose Fock space is 
built using creation operators $a^\dagger_1,\ldots,a^\dagger_{s_1}$ and 
$b^\dagger_1,\ldots,
b^\dagger_{s_2}$, acting on vacuum $\mid {\bf 0};{\bf 0}\rangle$. 
The $a_i$ and $b_i$ obey the CCR/CAR's. The basis 
elements are written\footnote{We note in passing that the basis used in 
Eq.~(2) of Ref.~\cite{Pol2} for fermionic systems is not properly
anticommuting.}
\begin{equation}
\mid {\bf m};{\bf n} \rangle = 
\prod_{i=1}^{s_1} (m_i!)^{-1/2} (a^\dagger_i)^{m_i}
\prod_{i=1}^{s_2} (n_i!)^{-1/2} (b^\dagger_i)^{n_i}
\mid {\bf 0};{\bf 0}\rangle,
\label{eq:Fock}
\end{equation}
and we write $m=\sum m_i$, $n=\sum n_i$ etc.
We will also need the alternative basis 
$\mid {\bf m};{\bf \tilde{n}}\rangle$ defined by 
\begin{equation}
\mid {\bf m};{\bf \tilde{n}} \rangle = 
\prod_{i=1}^{s_1} (m_i!)^{-1/2} (a^\dagger_i)^{m_i}
\prod_{i=1}^{s_2} (n_i!)^{-1/2} ((A^\dagger b)_i^\dagger)^{n_i}
\mid {\bf 0};{\bf 0}\rangle.
\end{equation}

Suppose the evolution $U$ on Fock space is unitary and such that
\begin{eqnarray}
a_i(T) & = & U^\dagger a_i U = P_{ij} a_j + Q_{ij} b_j  \nonumber \\
b_i(T) & = & U^\dagger b_i U = R_{ij} a_j + S_{ij} b_j ,
\label{eq:th}
\end{eqnarray}
where the matrix
\begin{equation}
V = \left(\begin{array}{cc} P & Q \\ R & S \end{array}\right)
\end{equation}
is unitary. We note that this entails that $U$ preserves the total
particle number $\sum a^\dagger_i a_i + \sum b^\dagger_i b_i$.

We now specialise to the bosonic case. From Section~\ref{sect:scpigf},
the evolution operator $X$ has matrix elements given by
\begin{equation}
\langle {\bf m};{\bf n} \mid X \mid {\bf m}^\prime;{\bf n}^\prime \rangle 
= {\cal N}_{\rm b} \delta_{{\bf nn}^\prime}
\sum_{{\bf n}''} 
\langle {\bf m};{\bf \tilde{n}''}\mid U\mid {\bf m}^\prime ;{\bf n}''\rangle ,
\label{eq:PI}
\end{equation}
where ${\cal N}_{\rm b}$ is a normalisation constant, chosen to ensure
that $\langle {\bf 0}\mid X\mid {\bf 0}\rangle = 1$ (as it should be
for any free theory). 
This allows us to evaluate ${\cal N}_{\rm b}$ explicitly,
because the matrix element $\langle {\bf 0};{\bf \tilde{n}} \mid U 
\mid{\bf 0}; {\bf n} \rangle$ is 
\begin{equation}
\langle {\bf 0};{\bf \tilde{n}} \mid U 
\mid{\bf 0}; {\bf n} \rangle =
\langle {\bf 0};{\bf 0} \mid \prod_i 
\frac{(A^\dagger b)_i^{n_i}}{(n_i!)^{1/2}} U 
\mid{\bf 0}; {\bf n} \rangle
= \langle {\bf 0};{\bf 0}\mid U \prod_i 
\frac{(A^\dagger b(T))_i^{n_i}}{(n_i!)^{1/2}}
\mid{\bf 0}; {\bf n} \rangle ,
\end{equation}
and is therefore equal to the coefficient of $\prod_i x_i^{n_i}$ in the
expansion of $\prod_i (\sum_j (A^\dagger S)_{ij}x_j)^{n_i}$. We have
used the fact that $U$ preserves the vacuum. The generating function 
$G(x_1,\ldots,x_{s_2})$ for these
coefficients can be found in \S 66 of~\cite{Mac}, and is given by
\begin{equation}
G(x_1,\ldots,x_{s_2}) = \frac{(-1)^{s_2} (x_1x_2\ldots
x_{s_2})^{-1}}{\det (A^\dagger S - {\rm diag}\, (x_1^{-1},x_2^{-1},
\ldots,x_{s_2}^{-1}))}. 
\end{equation}
The sum over all ${\bf n}$ of these
matrix elements is obtained simply by evaluating the generating function
with all $x_i$ equal to unity. Thus we obtain
\begin{equation}
{\cal N}_{\rm b} = \det (\openone - A^\dagger S).
\label{eq:N}
\end{equation}

Next, we claim that
\begin{equation}
X^{-1} a_i X =  M_{ij}a_j,
\label{eq:comm}
\end{equation}
where $M=P+Q(A-S)^{-1}R$ is unitary. Together with the trivial
evolution $X^{-1} b_i X =B_{ij}b_j$, this shows that $X$ is unitary. Moreover,
this is the free evolution derived in various ways in the body of the
paper. 

{}To establish~(\ref{eq:comm}), we first note that 
\begin{eqnarray}
\sum_{{\bf n}} 
\langle {\bf m};{\bf \tilde{n}} \mid  U b_i 
\mid {\bf m}^\prime; {\bf n} \rangle 
&=& \sum_{{\bf n}} 
\langle {\bf m};{\bf \tilde{n}} \mid (A^\dagger b)_i U 
\mid {\bf m}^\prime; {\bf n} \rangle 
\nonumber \\
&=& \sum_{{\bf n}} 
\langle {\bf m};{\bf \tilde{n}} \mid  U A^\dagger_{ij}
(R_{jk}a_k+S_{jk}b_k) \mid {\bf m}^\prime; {\bf n} \rangle ,
\label{eq:A7}
\end{eqnarray}
where the first step follows by relabelling the sum over $n_i$. 
Collecting terms in the $b_i$ and rearranging, we have
\begin{equation}
\sum_{{\bf n}} 
\langle {\bf m};{\bf \tilde{n}} \mid  U b_i 
\mid {\bf m}^\prime; {\bf n} \rangle 
= \sum_{{\bf n}} 
\langle {\bf m};{\bf \tilde{n}} \mid  U 
(A-S)^{-1}_{ij} R_{jk}a_k\mid {\bf m}^\prime; {\bf n} \rangle ,
\end{equation}
and hence 
\begin{eqnarray}
\sum_{{\bf n}} 
\langle {\bf m};{\bf \tilde{n}} \mid a_i U 
\mid {\bf m}^\prime; {\bf n} \rangle 
&=& \sum_{{\bf n}} 
\langle {\bf m};{\bf \tilde{n}} \mid U (P_{ij} a_j + Q_{ij} b_j) \mid
{\bf m}^\prime;{\bf n} \rangle \nonumber \\
&=& \sum_{{\bf n}} 
\langle {\bf m};{\bf \tilde{n}} \mid  U M_{ij}a_j 
\mid {\bf m}^\prime; {\bf n} \rangle ,
\end{eqnarray}
where $M=P+Q(A-S)^{-1}R$. Thus we have $a_i X = X M_{ij}a_j$ as required.

In the fermionic case, we define the operator $X$ by
\begin{equation}
\langle {\bf m};{\bf n} \mid X \mid {\bf m}^\prime;
{\bf n}^\prime \rangle = {\cal N}_{\rm f}\delta_{{\bf nn}^\prime}
\sum_{{\bf n}''} (-1)^{n''}
\langle {\bf m};{\bf \tilde{n}''} \mid U \mid {\bf m}^\prime ;{\bf n}'' 
\rangle ,
\label{eq:PIf}
\end{equation}
where ${\cal N}_{\rm f}$ is chosen to ensure
that $\langle {\bf 0}\mid X\mid {\bf 0}\rangle = 1$. The factor of 
$(-1)^{n''}$ is necessary in order to obtain agreement with the 
canonical theory. To see this, note that the first step in~(\ref{eq:A7})
is not valid in the fermionic case, due to the anticommutation relations
satisfied by the $a_i$ and $b_i$ and the definition~(\ref{eq:Fock}).
Instead, the corresponding result is
\begin{equation}
\sum_{{\bf n}} (-1)^{n(m+m^\prime)}
\langle {\bf m};{\bf \tilde{n}} \mid  U b_i \mid {\bf m}^\prime; {\bf n} 
\rangle  = \sum_{{\bf n}} (-1)^{n(m+m^\prime)}
\langle {\bf m};{\bf \tilde{n}} \mid (A^\dagger b)_i U 
\mid {\bf m}^\prime; {\bf n} \rangle  ,
\end{equation}
in which the factors of $(-1)^{m}$ and $(-1)^{m^\prime}$
arise from anticommuting $b_i$ past the string of creation operators for 
$\mid{\bf m}\rangle$ and $\mid{\bf m}^\prime\rangle$ respectively.
We may replace $(-1)^{n(m+m^\prime)}$ by  $(-1)^{n}$ because $U$
preserves the total particle number and therefore the summands can be
nonzero only when $m^\prime= m+1$. 
 
Exactly analogous arguments to those for the bosonic case then show
that eq.~(\ref{eq:comm}) holds, and that $X$ is unitary. Thus we have
obtained agreement with the canonical theory. 

The constant ${\cal N}_{\rm f}$ is easily evaluated once it has
been expressed in the form
\begin{equation}
{\cal N}_{\rm f}^{-1} = \langle {\bf 0};{\bf 0}\mid \left[
\bigwedge{}^{s_2} (\openone - A^\dagger S)b_{s_2}\ldots b_1 \right]
b_1^\dagger \ldots b_{s_2}^\dagger \mid {\bf 0};{\bf 0} \rangle ,
\label{eq:fclaim}
\end{equation}
for then one may use the exterior algebra definition of the 
determinant\footnote{Recall that if $Q$ is an endomorphism of a
vector space with anticommuting basis $b_1,\ldots,b_N$, then 
$(\bigwedge{}^N Q)b_N\ldots b_1 = (Qb)_N\ldots (Qb)_1 = 
(\det Q)b_N\ldots b_1$.} to conclude that
\begin{equation}
{\cal N}_{\rm f} = \left[ 
\det (\openone -A^\dagger S)\right]^{-1}.
\end{equation}
To establish Eq.~(\ref{eq:fclaim}), we write its RHS as ${\cal N}^{-1}$
and  expand the exterior power to obtain
\begin{equation}
{\cal N}^{-1} = \sum_{\bf n} \langle {\bf 0};{\bf 0}\mid 
(-1)^n c_{s_2}^{(n_{s_2})}\ldots c_1^{(n_1)} b_1^\dagger\ldots
b_{s_2}^\dagger \mid {\bf 0};{\bf 0}\rangle,
\end{equation}
where $c_i^{(n_i)}$ is defined to be equal to $b_i$ if $n_i=0$ or
$(A^\dagger S b)_i$ if $n_i=1$. Next, move the leftmost
$c_i^{(n_i)}$ with $n_i=0$ rightwards using the anticommutation relations
until it sits next to $b_i^\dagger$, at which point the $b_i b_i^\dagger$
combination may be removed by a further application of the CAR's.
Repeating the process until all $c_i^{(0)}$'s have been removed, 
one eventually finds
\begin{equation}
{\cal N}^{-1} = \sum_{\bf n} (-1)^n \langle {\bf 0};{\bf 0}\mid
(A^\dagger S b)_i^{n_i} \mid {\bf 0};{\bf n}\rangle ,
\end{equation}
which is easily shown to be equal to $\sum_{\bf n} (-1)^n
\langle {\bf 0};{\bf \tilde{n}}\mid U\mid {\bf 0};{\bf n}\rangle = 
{\cal N}_{\rm f}^{-1}$, thus verifying our claim.
 
\section{Violation of CCR/CAR's in the 3-Point Model}
\label{sect:detail}

We present here the details of the calculation leading to
Eqs.~(\ref{eq:J1}),~(\ref{eq:K1}) and~(\ref{eq:J2}) and the statements
made at the end
of Sect.~\ref{sect:fact}. We consider the 1-parameter family of operator  
orderings labelled by $\alpha\in[0,1]$ discussed in
Sect.~\ref{sect:fact} for which
\begin{equation}
\Psi(t) = e^{-i\lambda T\alpha\Psi(0)^\dagger\Psi(0)}
\left(e^{-iWT}\Psi(0)\right)
e^{-i\lambda T(1-\alpha)\Psi(0)^\dagger\Psi(0)} ,
\end{equation}
and consider Bose and Fermi statistics simultaneously, seeking
solutions of the form $b=f(d^\dagger d,c^\dagger c) c$ where 
\begin{equation}
c = \frac{R_1 a_1 +R_2a_2}{\|R\|} \qquad {\rm and}\qquad 
d=\frac{\overline{R}_2a_1-\overline{R}_1a_2}{\|R\|} 
\end{equation}
obey the same commutation relations as $a_1$ and $a_2$.

Applying $b$ to elements of form $(d^\dagger)^m c^\dagger\mid 0\rangle$, 
we obtain the consistency requirement
\begin{equation}
f(m,0) = \left(\|R\|+S f(m,0)\right) e^{-i\lambda
T[m+(1-\alpha)(1+|f(m,0)|^2)]},
\end{equation}
and applying $b$ to elements of form $(d^\dagger)^m
(c^\dagger)^{n+2}\mid 0\rangle$ for $m,n\ge 0$, we obtain the recursion 
relation 
\begin{equation}
f(m,n+1) = \left(\|R\|+S f(m,n+1)\right) e^{-i\lambda
T[m+\alpha(n+1)(1+|f(m,n)|^2)+(1-\alpha)(n+2)(1+|f(m,n+1)|^2)]} .
\end{equation}

We compute the quantity
\begin{equation}
J = \langle 0\mid (a_2(T)a_1(T)\mp a_1(T)a_2(T))d^\dagger c^\dagger\mid
0\rangle
\end{equation}
for Bose ($-$) and Fermi ($+$) statistics and also
\begin{equation}
K= \langle 0\mid a_1(T)^2 d^\dagger c^\dagger\mid 0\rangle 
\end{equation}
for Fermi statistics. Note that $d^\dagger c^\dagger \mid 0\rangle$
is an element of ${\frak F}_2$; in the fermionic case it is $a_1^\dagger
a_2^\dagger\mid 0\rangle$.

First note that
\begin{equation}
\langle 0\mid a_i(T)a_j(T) d^\dagger c^\dagger\mid 0\rangle = 
e^{-i\omega}\langle 0\mid (P_{ik}a_k+Q_i b)e^{-i\lambda
Tc^\dagger |f(d^\dagger d, c^\dagger c)|^2c} (P_{jl}a_l+Q_j b) 
d^\dagger c^\dagger\mid 0 \rangle ,
\end{equation} 
where $\omega = \lambda
T\left[(1-\alpha)(2+|f(1,0)|^2)+1\right]$. We have
\begin{eqnarray}
P_{1i}a_i &=& \frac{P_{1i}\overline{R}_i}{\|R\|} c+
\frac{P_{11}R_2-P_{12}R_1}{\|R\|} d \nonumber \\
&=& -\frac{Q_1 \overline{S}}{\|R\|} c - \frac{\overline{Q}_2\det
P}{\overline{S} \|R\|} d ,
\end{eqnarray}
where we have used the identities
\begin{equation}
P_{ij}\overline{R}_j = -Q_i \overline{S} ,
\end{equation}
and
\begin{equation}
P_{11}R_2-P_{12}R_1 = -\frac{\overline{Q}_2\det P}{\overline{S}} ,
\end{equation}
which follow from the unitarity of $e^{-iWT}$. The second of these
is derived by noting that $\overline{S}R_j = -\overline{Q}_i P_{ij}$.
Similarly, we have
\begin{equation}
P_{2i}a_i = -\frac{Q_2 \overline{S}}{\|R\|} c + \frac{\overline{Q}_1
\det P}{\overline{S} \|R\|} d ,
\end{equation}
in which we have used $P_{21}R_2-P_{22}R_1=\overline{Q_1}\det
P/\overline{S}$. 

We can now compute
\begin{eqnarray}
\langle 0 \mid a_2(T)a_1(T) d^\dagger c^\dagger\mid 0\rangle &= &
e^{-i\omega}\frac{\det P}{\overline{S}\|R\|}
\left\{\pm |Q_1|^2\left(f(1,0)-\frac{\overline{S}}{\|R\|}\right)-\right.
\nonumber \\
& & \left.
|Q_2|^2\left(f(0,0)-\frac{\overline{S}}{\|R\|}\right)e^{-i\lambda 
T|f(0,0)|^2}\right\} ,
\end{eqnarray}
where the $\pm$ is $+$ for Bose and $-$ for Fermi statistics. To obtain
$J$, we interchange the suffices $1$ and $2$ and {\em add} for
Bose and {\em subtract} for Fermi. The sign reversal occurs because
$d^\dagger c^\dagger \mid 0\rangle$ flips sign under the
interchange. The final result is then
\begin{equation}
J= 
e^{-i\omega} \det P(\pm |Q_1|^2+|Q_2|^2)\left\{
\frac{f(1,0) - f(0,0)e^{-i\lambda T|f(0,0)|^2}}{\overline{S}\|R\|}
+\frac{e^{-i\lambda T |f(0,0)|^2}-1}{\|R\|^2} 
\right\} .
\label{eq:J}
\end{equation}

The matrix element $K$ may be computed for Fermi statistics as
\begin{equation}
K=e^{-i\omega} Q_1\overline{Q}_2\det P
\left\{
\frac{f(1,0) - f(0,0)e^{-i\lambda T|f(0,0)|^2}}{\overline{S}\|R\|}
+\frac{e^{-i\lambda T |f(0,0)|^2}-1}{\|R\|^2}\right\} .
\label{eq:K}
\end{equation}

With the particular operator ordering used in Section~\ref{sect:inter}
($\alpha=1$), we have $f(0,0) =\|R\|(1-S)^{-1}$ and 
$f(1,0)=\|R\|(e^{i\lambda T}-S)^{-1}$. Substituting these values into
Eqs.~(\ref{eq:J}) and~(\ref{eq:K}) and using the fact that
$\|R\|^2=1-|S|^2$, we obtain Eqs.~(\ref{eq:J1}),~(\ref{eq:K1})
and~(\ref{eq:J2}) respectively.  

Finally, one should also check that the expression enclosed within braces
in Eqs.~(\ref{eq:J}) and~(\ref{eq:K}) does not vanish. For $\lambda T\ll
1$, one may prove this by perturbing about the free solution to obtain
$f(0,0)$ and $f(1,0)$ to second order in $\lambda T$ if $S\not\in{\Bbb
R}$. If $S$ is real, one needs to go to third order. 

\narrowtext

\widetext

\mediumtext
\begin{table}
\caption{Table of $\theta_n$ defined by Eq.~(\protect\ref{eq:thndf}) for
different
values of the coupling strength $\mu$. The final row shows the value of 
$\theta$ corresponding to the unique classical limit picked out by the
quantum theory.}
\begin{tabular}{rdddd} 
$n$ & \multicolumn{4}{c}{$\theta_n$} \\
        & \multicolumn{1}{r}{$\mu=0.5$} & 
\multicolumn{1}{r}{$\mu=3.0$} & \multicolumn{1}{r}{$\mu=7.0$} & 
\multicolumn{1}{r}{$\mu=13.0$} \\ \hline
       500 & 0.9020    & $-$2.7496 & 2.2226    & 2.7558 \\
      1000 & 0.9018    & $-$2.7492 & 2.2124    & 2.7527 \\
      1500 & 0.9017    & $-$2.7490 & 2.2115    & 2.7516 \\
      2000 & 0.9016    & $-$2.7489 & 2.2112    & 2.7510 \\
      2500 & 0.9016    & $-$2.7489 & 2.2109    & 2.7507 \\
\hline
Classical 
solution   & 0.9015    & $-$2.7487   & 2.2101    & 2.7494 \\
\end{tabular}

\label{table1}
\end{table}
\narrowtext

\end{document}